\documentclass[twocol]{ametsoc_navid}

\journal{jas}

\usepackage{ulem}

\usepackage{comment}
\usepackage{enumerate}

\bibpunct{(}{)}{;}{a}{}{,}

\newcommand\xoutpars[1]{\let\helpcmd\xout\parhelp#1\par\relax\relax}
\newcommand\soutpars[1]{\let\helpcmd\sout\parhelp#1\par\relax\relax}
\long\def\parhelp#1\par#2\relax{%
  \helpcmd{#1}\ifx\relax#2\else\par\parhelp#2\relax\fi%
}

\graphicspath{ {Figures/} }

\ifdraft
\else
\usepackage{doi}
\usepackage{hyperref}
\hypersetup{
	breaklinks=true,
	colorlinks=true,
	linkcolor=blue,
	citecolor=Purple,
	pdfstartview={FitH},
	pdfview={FitH 0},
	pdfauthor={N A Bakas, N C Constantinou and P J Ioannou},
	pdftitle={Statistical state dynamics of weak jets in barotropic beta-plane turbulence},
 }
\fi

\usepackage{cancel}

\usepackage{scalerel}
\usepackage[usestackEOL]{stackengine}
\let\oldmathchoice\mathchoice
\usepackage{breqn}
\let\newmathchoice\mathchoice

\def\dashint{\let\mathchoice\oldmathchoice\,\ThisStyle{\ensurestackMath{%
            \stackinset{c}{.2\LMpt}{c}{.5\LMpt}{\SavedStyle-}{%
            \SavedStyle\phantom{\int}}}%
        \setbox0=\hbox{$\SavedStyle\int\,$}\kern-\wd0}\int%
        \let\mathchoice\newmathchoice}

\def\deg{$^{\circ}$N}

\newcounter{saveeqn}%

\newcommand{\bdm}{\begin{equation*}}
\newcommand{\edm}{\end{equation*}}
\newcommand{\bea}{\begin{eqnarray}}
\newcommand{\eea}{\end{eqnarray}}

\newcommand{\partialf}[2]
{
 \ifthenelse{\equal{#1}{}}{\frac{\partial}{\partial #2}}{\frac{\partial #1}{\partial #2}}
}

\newcommand{\real}{\mathop{\mathrm{Re}}}

\renewcommand{\(}{\left(}
\renewcommand{\)}{\right)}
\renewcommand{\[}{\left[}
\renewcommand{\]}{\right]}
\newcommand{\<}{\left\langle}
\renewcommand{\>}{\right\rangle}

\newcommand{\Del}{\Delta}

\renewcommand{\d}{\delta}

\newcommand{\df}{d}

\newcommand{\s}{\sigma}
\renewcommand{\b}{\beta}

\renewcommand{\v}{\upsilon}
\newcommand{\z}{\zeta}

\renewcommand{\i}{\mathrm{i}}

\providecommand\bcdot{\boldsymbol{\cdot}}

\newsavebox{\astrutbox}
\sbox{\astrutbox}{\rule[-5pt]{0pt}{20pt}}

\newcommand\p{\ensuremath{\partial}}

\newcommand{\e}{\varepsilon}
\newcommand{\ec}{\varepsilon_c}

\def\bit{\vphantom{\dot{W}}}

\def\thet{\vartheta}

\def\sR{s_r}
\def\sR{f_r}

\def\Fcal{\mathcal{F}}
\def\Acal{\mathcal{A}}

\def\Lcal{\mathcal{L}}
\def\Rcal{\mathcal{R}}

\def\Ncal{\mathcal{N}}
\def\Pcal{\mathcal{P}}
\def\deg{^\circ}

\def\bx{\boldsymbol{x}}

\def\bu{\boldsymbol{u}}

\def\nv{\boldsymbol{n}}
\def\kv{\boldsymbol{k}}

\def\fr{\sR}

\def\nablav{\bm\nabla}

\newcommand{\defn}{\ensuremath{\stackrel{\mathrm{def}}{=}}}
\renewcommand{\equiv}{\defn}
\newcommand{\ee}{\mathrm{e}}
\renewcommand{\dashint}{\int_\infty}

\newcommand{\um}{\bar{u}}

\newcommand{\zd}{\zeta_{*}}
\newcommand{\psid}{\psi_{*}}
\newcommand{\bud}{\bu_{*}}
\newcommand{\ud}{u_{*}}
\newcommand{\vd}{\v_{*}}
\newcommand{\betad}{\beta_{*}}
\newcommand{\rd}{r_{*}}
\newcommand{\bxd}{\bx_{*}}
\newcommand{\xd}{x_{*}}
\newcommand{\yd}{y_{*}}
\newcommand{\td}{t_{*}}

\newcommand{\Ld}{L_{*}}
\newcommand{\kvd}{\kv_{*}}
\newcommand{\kd}{k_{*}}
\newcommand{\kxd}{k_{x*}}
\newcommand{\kyd}{k_{y*}}
\newcommand{\kfd}{k_{f*}}
\newcommand{\dfd}{\delta_{f*}}
\newcommand{\ed}{\varepsilon_{*}}

\newcommand{\xid}{\xi_{*}}
\newcommand{\ncd}{n_{c*}}
\newcommand{\Qd}{Q_{*}}

\newcommand{\Qhatd}{\hat{Q}_{*}}
\newcommand{\nablavd}{\nablav_{*}}
\newcommand{\Deld}{\Del_{*}}

\newcommand{\uhato}{\hat{\bar{u}}_{1}}
\newcommand{\uhatt}{\hat{\bar{u}}_{2}}
\newcommand{\Chat}{\hat{\bar{C}}}

\title{Statistical state dynamics of weak jets in barotropic beta-plane turbulence}

\authors{Nikolaos A.  Bakas}
 \affiliation{Laboratory of Meteorology and Climatology, Department of Physics, University of Ioannina, Ioannina, Greece}

\extraauthor{Navid C.  Constantinou\correspondingauthor{Navid Constantinou, 142 Mills Road, Building Jaeger 8, Research School of Earth Sciences, Australian National University, Canberra, Australian Capital Territory, Australia.}}
\extraaffil{\ifdraft\else\footnotesize\fi Scripps Institution of Oceanography, University of California San Diego, La Jolla, California, USA\\
Research School of Earth Sciences, Australian National University, Canberra, Australian Capital Territory, Australia\\
ARC Centre of Excellence for Climate Extremes, Australian National University, Canberra, ACT, 2601, Australia}

\extraauthor{Petros J.  Ioannou}
\extraaffil{\ifdraft\else\footnotesize\fi Department of Physics, National and Kapodistrian University of Athens, Athens, Greece}

\email{navid.constantinou@anu.edu.au}

\abstract{
Zonal jets in a barotropic setup emerge out of homogeneous turbulence through a flow-forming instability of the homogeneous turbulent state (`zonostrophic instability') which occurs as the turbulence intensity increases.  This has been demonstrated using the statistical state dynamics (SSD) framework with a closure at second order.  Furthermore, it was shown that for small supercriticality the flow-forming instability follows Ginzburg--Landau (\mbox{G--L}) dynamics.  Here, the SSD framework is used to study the equilibration of this flow-forming instability for small supercriticality.  First, we compare the predictions of the weakly nonlinear \mbox{G--L} dynamics to the fully nonlinear SSD dynamics closed at second order for a wide ranges of parameters.  A new branch of jet equilibria is revealed that is not contiguously connected with the \mbox{G--L} branch.  This new branch at weak supercriticalities involves jets with larger amplitude compared to the ones of the \mbox{G--L} branch.  Furthermore, this new branch continues even for subcritical values with respect to the linear flow-forming instability.  Thus, a new \emph{nonlinear} flow-forming instability out of homogeneous turbulence is revealed.  Second, we investigate how both the linear flow-forming instability and the novel nonlinear flow-forming instability are equilibrated.  We identify the physical processes underlying the jet equilibration as well as the types of eddies that contribute in each process.  Third, we propose a modification of the diffusion coefficient of the \mbox{G--L} dynamics that is able to capture the evolution of weak jets at scales other than the marginal scale (side-band instabilities) for the linear flow-forming instability.}

\begin{document}

\maketitle


\section{Introduction\label{sec:intro}}

Robust eddy-driven zonal jets are ubiquitous in planetary atmospheres \citep{Ingersoll-90,Ingersoll-etal-2004,Vasavada-and-Showman-05}.  Laboratory experiments, theoretical studies, and numerical simulations show that small-scale turbulence self-organizes into large-scale coherent structures, which are predominantly zonal and, furthermore, that the small-scale turbulence supports the jets against eddy mixing \citep{Starr-1968, Huang-Robinson-98, Read-etal-2007, Salyk-etal-2006}. One of the simplest models, which is a testbed for theories regarding turbulence self-organization, is forced--dissipative barotropic turbulence on a beta-plane.

An advantageous framework for understanding coherent zonal jet self-organization is the study of the Statistical State Dynamics (SSD) of the flow.  SSD refers to the dynamics that governs the statistics of the flow rather than the dynamics of individual flow realizations.  However, evolving the hierarchy of the flow statistics of a nonlinear dynamics soon becomes intractable; a turbulence closure is needed.  Unlike the usual paradigm of homogeneous isotropic turbulence, when strong coherent flows coexist with the incoherent turbulent field, the SSD of the turbulent flow is well captured by a second-order closure \citep{Farrell-Ioannou-2003-structural,Farrell-Ioannou-2007-structure,Farrell-Ioannou-2009-equatorial,Tobias-etal-2011, Srinivasan-Young-2012, Bakas-Ioannou-2013-prl,Tobias-Marston-2013, Constantinou-etal-2014,Constantinou-etal-Madrid-2014, Thomas-etal-2014, Ait-Chaalal-etal-2016, Constantinou-etal-2016,Farrell-etal-2016-VLSM, Farrell-Ioannou-2017-Saturn,Fitzgerald-Farrell-2018, Frishman-Herbert-2018}. Such a second-order closure comes in the literature under two names: `S3T',  which stands for Stochastic Structural Stability Theory \citep{Farrell-Ioannou-2003-structural} and `CE2', which stands for Cumulant Expansion of second order \citep{Marston-etal-2008}.  Hereafter, we refer to this second-order closure as~S3T.

Using the S3T second-order closure it was first theoretically predicted that zonal jets in barotropic beta-plane turbulence emerge spontaneously out of a background of homogeneous turbulence through an \emph{instability of the SSD} \citep{Farrell-Ioannou-2007-structure, Srinivasan-Young-2012}.  That is, S3T predicts that jet formation is a bifurcation phenomenon, similar to phase transitions, that appears as the turbulence intensity crosses a critical threshold.  This prediction comes in contrast with the usual theories for zonal jet formation that involve anisotropic arrest of the inverse energy cascade at the Rhines' scale \citep{Rhines-1975,Vallis-Maltrud-93}. Jet emergence as a bifurcation was subsequently confirmed by comparison of the analytic predictions of the S3T closure with direct numerical simulations \citep{Constantinou-etal-2014,Bakas-Ioannou-2014-jfm}. This flow-forming SSD instability is markedly different from hydrodynamic instability in which the perturbations grow in a fixed mean flow. In the flow-forming instability, both the coherent mean flow and the incoherent eddy field are allowed to change. The instability manifests as follows: a weak zonal flow that is inserted in an otherwise homogeneous turbulent field, organizes the incoherent fluctuations to \emph{coherently} reinforce the zonal flow. This instability has analytic expression only in the SSD and we therefore refer to this new kind of instabilities as `SSD instabilities'.  In particular, the flow-forming `SSD instability' of the homogeneous turbulent state to zonal jet mean flow perturbations is also referred to as `zonostrophic instability' \citep{Srinivasan-Young-2012}.

\begin{figure}[t]
\centerline{\includegraphics[width=16pc]{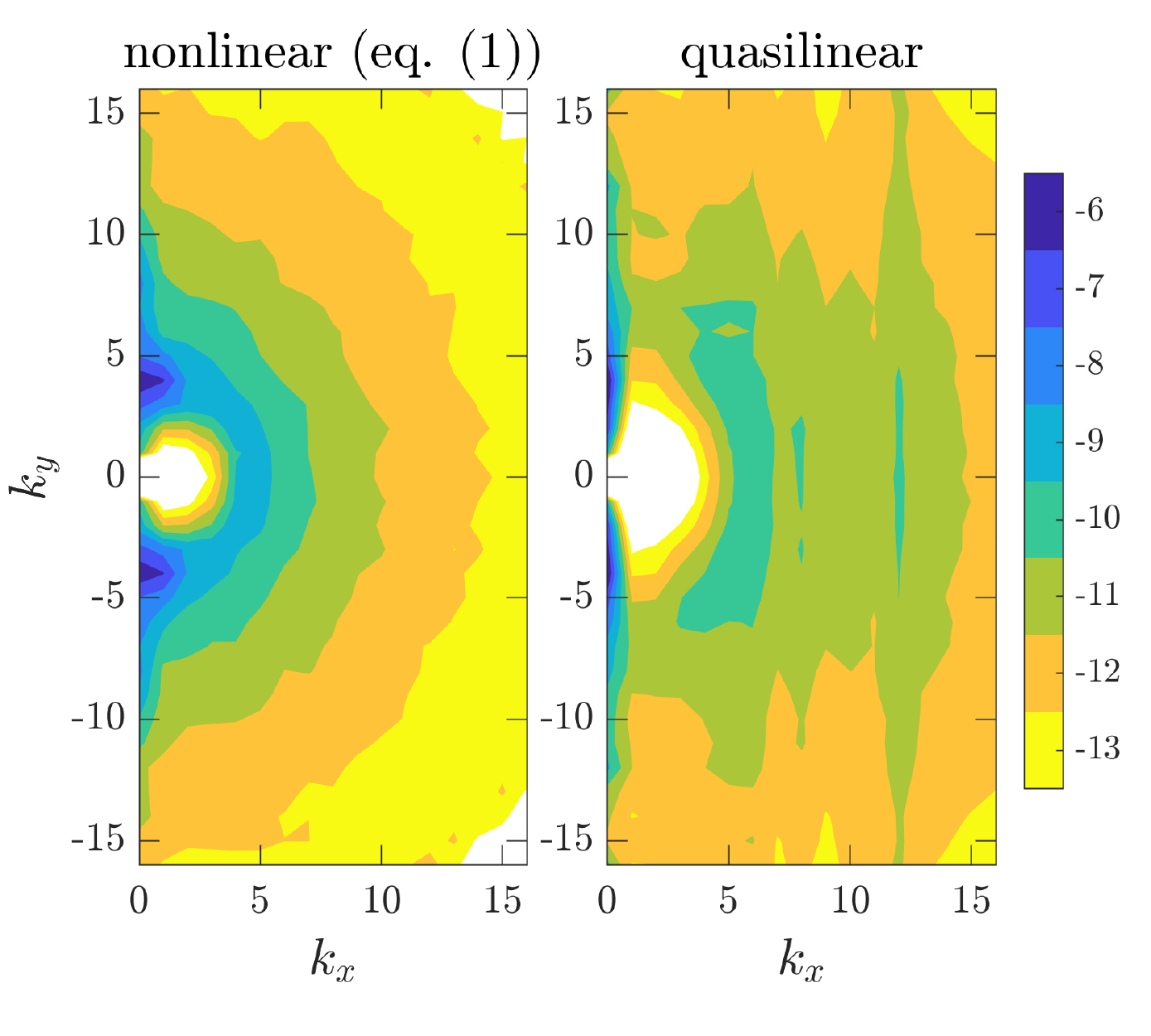}}
\vspace{-1em}
\caption{Second-order closure can captures the mean flow dynamics despite differences in structure of eddy spectra. Here shown are the energy spectra for a fully nonlinear simulation (eq.~\eqref{eq:NLbarotropic}) and its quasilinear approximation (i.e., employing the second-order closure). Both simulations form 4 strong jets of similar strength. Setup as described in section~\ref{sec:compare} with $\beta/(k_f r) = 70$ and $\e/(k_f^2 r^3) = 4\times 10^5$. Contours in logarithmic scale and the same for both panels.} \label{fig:spectra}
\end{figure}

\citet{Kraichnan-1976} suggested that the large-scale mean flow is supported by small-scale eddies. Indeed, when the large scales dominate the eddy field (i.e., when the large-scale shear time, $\tau_m$, is far shorter than the eddy turnover time, $\tau_e$) the small-scale eddies have the tendency to flux momentum and support large-scale mean flows \citep{Shepherd-1987-jfm,Huang-Robinson-98,Chen-etal-2006,Holloway-10,Frishman-Herbert-2018}.  Under such circumstances, we expect the S3T second-order closure of the SSD to be accurate. Furthermore, \citet{Bouchet-etal-2013} provided a  proof that in the limit  $\tau_e/\tau_m \to \infty$ the SSD of large-scale jets in equilibrium with their eddy field are governed exactly by a second-order closure. Recent studies revealed that the second-order closure remains accurate even at moderate scale separation between $\tau_m$ and $\tau_e$ (see, e.g.,~\cite{Srinivasan-Young-2012,Marston-etal-2014,Marston-etal-2016,Frishman-etal-2017,Frishman-Herbert-2018}). That is, the second-order closure manages to reproduces fairly accurately the structure of the mean flow \emph{even though} there could be differences in the eddy spectra and the concomitant eddy correlations; see, e.g., figure~\ref{fig:spectra}.

However, surprisingly enough, \emph{S3T remains accurate even at a perturbative level}, i.e.,~when the mean flows/jets are just emerging with $\tau_e/\tau_m \to 0$ (the exactly opposite limit of \cite{Bouchet-etal-2013}). This perturbative-level agreement is reported by \cite{Constantinou-etal-2014,Bakas-Ioannou-2013-prl, Bakas-Ioannou-2014-jfm} for barotropic flows, by~\cite{Bakas-Ioannou-2018} for baroclinic flows, by~\citet{Fitzgerald-Farrell-2018} for vertically sheared stratified flows, by \cite{Constantinou-Parker-2018} for magnetized flows in astrophysical settings, and by~\cite{Farrell-Ioannou-2017-bifur} for the formation of spanwise varying mean flows and mean vortices (streaks--rolls) in 3D channel flows. The reason that the S3T second-order closure works well \emph{even for very weak  mean flows} should be attributed to the existence of the collective flow-forming instability which seems to overpower the disruptive eddy--eddy nonlinear interactions, as long as the turbulent intensity is not exceptionally strong (which in most physical situations is usually the case). 

The dynamics that underlie the flow-forming SSD instability of the homogeneous state is well understood; \citet{Bakas-Ioannou-2013-jas} and \cite{Bakas-etal-2015} studied in detail this eddy--mean flow dynamics for barotropic flows and \citet{Fitzgerald-Farrell-2018-jas} for stratified flows. In these studies, the structures of the eddy field that produce up-gradient momentum fluxes, and thus drive the instability, were determined in the appropriate limit $\tau_{\rm diss}/\tau_m\to0$, with $\tau_{\rm diss}$ the dissipation time-scale.  



While the processes by which the flow-forming instability manifests are well understood, we lack comprehensive understanding of how this instability is equilibrated. For example, as the zonal jets grow they often merge or branch to larger or smaller scales \citep{Danilov-04,Manfroi-Young-99}, multiple turbulence--jet equilibria exist \citep{Farrell-Ioannou-2007-structure,Parker-Krommes-2013,Constantinou-etal-2014}, and, also, transitions from various turbulent jet attractors may occur \citep{Bouchet-etal-2018-jettransisions}.  Some outstanding questions include:
\begin{enumerate}[(\it i)]
	\item How is the equilibration of the flow-forming instability achieved and at which amplitude for the given parameters?
	\item What are the eddy--mean flow dynamics involved in the equilibration process as well as which eddies support the finite amplitude jets?
	\item
	What type of instabilities are involved in the observed jet variability phenomenology (jet merging and branching, multiple jet equilibria, transitions between various jet attractors) and what are the eddy--mean flow dynamics involved?
\end{enumerate}

To tackle these questions, \citet{Parker-Krommes-2013} first pointed out the analogy of jet formation and pattern formation
\citep{Hoyle-2006,Cross-Greenside-2009}.  Exploiting this analogy \citet{Parker-Krommes-2014-generation} were able to borrow tools and methods from pattern formation theory to elucidate the equilibration process.  In particular, they demonstrated that at small supercriticality, that is when the turbulence intensity is just above the critical threshold for jet formation, the nonlinear evolution of the zonal jets follows Ginzburg--Landau (\mbox{G--L}) dynamics.  In addition, \citet{Parker-Krommes-2014-generation} examined the quantitative accuracy of the \mbox{G--L} approximation by comparison with turbulent jet equilibria obtained from the fully nonlinear S3T dynamics.  Having established the validity of S3T dynamics even in the limit of very weak mean flows/jets (as we have discussed above), it is natural to then proceed studying the \mbox{G--L} dynamics of this flow-forming instability and its associated equilibration process. The perturbative-level agreement of the S3T predictions with direct numerical simulations of the full nonlinear dynamics argues that the study of the equilibration of the flow-forming instability using the \mbox{G--L} dynamics is well founded.

In this work, we revisit the small-supercriticality regime of \citet{Parker-Krommes-2014-generation}. We thoroughly test the validity
of the \mbox{G--L} approximation through a comparison with the fully nonlinear SSD closed at second order for a wide range parameter values (section~\ref{sec:compare}). Apart from the equilibrated flow-forming instability of the homogeneous turbulent state, which is governed by the \mbox{G--L} dynamics, we discover that an additional branch of jet equilibria exists for large values of $\beta/(k_f r)$ ($\beta$ is the planetary vorticity gradient, $r=1/\tau_{\rm diss}$ is the linear dissipation rate, and $1/k_f$ is the length scale of the forcing). This new branch of equilibria reveals that jets emerge as a cusp bifurcation, which implies that for large  $\beta/(k_f r)$ the emergent jets result from a nonlinear instability (see Fig~\ref{fig:bifurc_b58_jetspectra}(a)).
%

We investigate the eddy--mean flow dynamics involved in the equilibration of the flow-forming instabilities, as well as those involved in the secondary side-band jet instabilities that occur (section~\ref{sec:equil_landau}).  To do this, we derive the \mbox{G--L} equation in a physically intuitive way that allows for the comprehensive understanding of the nonlinear Landau term involved in the \mbox{G--L} equation (section~\ref{sec:equil}).
Using methods similar to the ones developed by \citet{Bakas-Ioannou-2013-jas} and~\cite{Bakas-etal-2015} we study the contribution of the forced eddies and their interactions in supporting the equilibrated finite amplitude jets (section~\ref{sec:equil_landau}). Finally, to elucidate the equilibration of the new branch of jet equilibria that are not governed by the \mbox{G--L} dynamics, we develop an alternative reduced dynamical system which generalizes the \mbox{G--L} equation  (section~\ref{sec:equil_landau}\ref{sec:upper_branch}). Using this reduced system we study the physical processes responsible for the equilibration of the new branch of jet equilibria.

%

\def\commentcolor{Purple}

\section{Statistical state dynamics of barotropic $\beta$-plane turbulence in the S3T second-order closure \label{sec:SSD}}

Consider a non-divergent flow $\bud=(\ud,\vd)$ on a $\beta$-plane with coordinates $\bxd=(\xd,\yd)$; $\xd$ is the zonal direction and $\yd$ the meridional direction.  Subscript asterisks here denote dimensional variables.  The flow is in an unbounded domain, unless otherwise indicated.  The flow is derived from a streamfunction $\psid$ via $(\ud,\vd) = (-\p_{\yd}\psid,\p_{\xd}\psid)$.  The relative vorticity of the flow is $\zd\equiv\p_{\xd}\vd-\p_{\yd}\ud=\Deld\psid$, with $\Deld\equiv\p^2_{\xd}+\p_{\yd}^2$ the Laplacian.  With stochastic excitation and linear dissipation the relative vorticity evolves according to:
\begin{equation}
(\partial_{\td}+\bud\bcdot\nablavd)(\zd+ \betad\yd)  = -\rd\zd + \sqrt{\ed}\xid\ .\label{eq:NLbarotropic}
\end{equation}
Linear dissipation at the rate $\rd$ parametrizes Ekman drag at the surface of the planet.  Turbulence is supported by the random stirring $\xid(\bxd,\td)$ that injects energy in the flow at rate~$\ed$.  This random stirring models vorticity sources such as convection and/or baroclinic growth processes that are absent in barotropic dynamics.  The random process $\xid$ is assumed \textit{(i)}~to have zero mean, \textit{(ii)}~to be spatially and temporally statistically homogeneous, and~\textit{(iii)}~to be
temporally delta-correlated  but spatially correlated.  Thus it satisfies:
\ifdraft\begin{align}
\<\xid(\bxd,\td)\>=0\quad\text{and}\quad\<\xid(\bx_{a*}, {\td}_1)\xi(\bx_{b*}, {\td}_2)\>=\Qd(\bx_{a*}-\bx_{b*})\,\d(t_{1*}-t_{2*})\ ,
\end{align}\else\begin{subequations}
\begin{align}
\<\xid({\bxd},{\td})\>&=0\quad\text{and}\\
\<\xid(\bx_{a*}, {\td}_1)\xid(\bx_{b*}, {\td}_2)\>&=\Qd(\bx_{a*}-\bx_{b*})\,\d(t_{1*}-t_{2*})\ ,
\end{align}\end{subequations}\fi
with $\Qd$ the homogeneous spatial covariance of the forcing.  Angle brackets denote ensemble averaging over realizations of the forcing.  The forcing covariance is constructed  by specifying a non-negative spectral power function $\Qhatd(\kvd)$ as:
\begin{equation}
\Qd(\bx_{a*}-\bx_{b*}) = \int \frac{\df^2 \kvd}{(2\pi)^2}\, \Qhatd(\kvd)\,\ee^{\i \kvd\bcdot (\bx_{a*}-\bx_{b*})}\label{eq:Qf}\ .
\end{equation}
In this work, we consider isotropic forcing with spectrum:
\begin{equation}
\Qhatd(\kvd) = 4\pi\,\kfd\,\d(\kd-\kfd)\ ,\label{eq:Qhat}
\end{equation}
where $\kd\defn|\kvd|$.  The forcing~\eqref{eq:Qhat} excites equally all waves with total wavenumber $\kfd$.  The forcing spectrum is normalized so that the total energy injection  is $\ed$.\footnote{In numerical simulations, we approximate the delta-function in~\eqref{eq:Qhat} as a gaussian with narrow width---see section~\ref{sec:compare} for more details.} 

Equation~\eqref{eq:NLbarotropic} is non-dimensionalized using the forcing length scale $\kfd^{-1}$ and the dissipation time scale $\rd^{-1}$.
The non-dimensional variables are: $\z=\zd/\rd$, $\bu=\bud/(k_{f*}^{-1}\rd)$, $\xi=\xid/(\kfd\sqrt{\rd})$, $\e =\ed /(k_{f*}^{-2} r_*^3)$,
$\b=\betad/(\kfd\rd)$ and $r=1$.  Thus, the non-dimensional version of~\eqref{eq:NLbarotropic} lacks all asterisks and has $r=1$.  The non-dimensional form of~$\Qhatd$ in~\eqref{eq:Qhat} is obtained dropping the asterisks and replacing $\kfd\mapsto1$.

The statistical state dynamics (SSD) of zonal jet formation in the S3T second-order closure comprise the dynamics of the first cumulant of the vorticity field $\bar\zeta(\bx,t)$, and of the second cumulant $C(\bx_a, \bx_b, t)\equiv \overline{\zeta'(\bx_a,t) \zeta'(\bx_b,t)}$.

The overbars here denote zonal average, while dashes denote fluctuations about the mean.  Thus, $\bar{\z} = -\partial_y\bar{u}$ and, the first cumulant of the flow can be equivalently described with $\bar{u}$.  Also, the eddy covariance $C$ is therefore homogeneous in $x$: $C(x_a-x_b, y_a, y_b, t)$.  Furthermore, the zonal average is assumed to satisfy the ergodic property, i.e.,~that the average of any quantity is equal to an ensemble average over realizations of $\xi$:~$\overline{\(\,\bcdot\, \)} = \<\( \,\bcdot\, \)\>$.

After dropping terms involving the third cumulant we can form the closed system for the evolution of the first and second cumulants of the flow:
	\begin{subequations}
	\label{eq:s3ts}
	\begin{align}
		\partial_t \bar u&=\Rcal(C)-\bar u\ \label{eq:s3tm} ,\\
		\partial_t C&=-\Lcal C+\Ncal(\bar u,C)+\varepsilon\,Q\ .\label{eq:cov_evo}
	\end{align}
	\end{subequations}
The derivation of~\eqref{eq:s3ts} has been presented many times; the reader is referred
to, e..g, the work by~\cite{Farrell-Ioannou-2003-structural,Srinivasan-Young-2012,Bakas-etal-2015}.  In~\eqref{eq:s3ts}, $\Lcal$ is the operator given in~\eqref{eq:op_L} that governs the linear eddy dynamics, and $\Ncal$ is the nonlinear operator {given in~\eqref{eq:op_N} that governs the interaction between the eddies and the instantaneous mean flow $\um(y,t)$.
The mean flow $\um$ is driven by the ensemble mean eddy vorticity flux $\overline{\v'\zeta'}$, which is expressed as a
linear function of the eddy vorticity covariance $C$ through $\Rcal(C)$ with $\Rcal$ given in~\eqref{eq:rs}.

The mean flow energy density, $E_m$, and the eddy energy density, $E_p$, are:
\ifdraft\begin{subequations}\begin{align}
  E_m &= \dashint \df^2\bx \;\frac1{2}\um^2\ ,\\
  E_p & = \dashint \df^2\bx \;\frac1{2} \[\bit\right.\overline{\bu'(\bx_a)\bcdot\bu'(\bx_b)}\left.\bit\]_{\bx_a=\bx_b} =  -\dashint \df^2\bx \;\frac1{4}\left[(\Del_a^{-1}+\Del_b^{-1}) C \right]_{\bx_a=\bx_b} \ .
\end{align}\label{eq:10}\end{subequations}\else
\begin{subequations}\begin{align}
  E_m &= \dashint \df^2\bx \;\frac1{2}\um^2\ ,\\
  E_p & = \dashint \df^2\bx \;\frac1{2} \[\bit\right.\overline{\bu'(\bx_a)\bcdot\bu'(\bx_b)}\left.\bit\]_{a=b}\nonumber\\
  &=  -\dashint \df^2\bx \;\frac1{4}\left[(\Del_a^{-1}+\Del_b^{-1}) C \right]_{a=b} \ ,
\end{align}\label{eq:10}\end{subequations} \fi
where $\dashint\equiv\lim_{L\rightarrow\infty}(2L)^{-2} \int_{-L}^{L}\int_{-L}^{L}$,
the subscripts on the Laplacian indicate the specific variable the operator is acting, and subscript $a=b$ implies that the function of $\bx_a$ and~$\bx_b$, e.g., inside the square brackets on the right-hand-side of~\eqref{eq:10}, is transformed into a function of a single variable by setting $\bx_a=\bx_b=\bx$.  The total averaged energy density relaxes over the dissipation scale (which is of $O(1)$ in the non-dimensional
equations) to the energy supported under stochastic forcing and dissipation:
	\begin{align}
		E(t)&\equiv E_m(t)+E_p(t)=\[E(0)-\frac{\e}{2}\]\ee^{-2t}+\frac{\e}{2}\ .\label{eq:tot_en}
	\end{align}
Therefore, the total energy remains bounded under S3T dynamics~\citep{Bakas-Ioannou-2015-book}.

\section{The flow-forming instability and the underlying eddy--mean flow dynamics}

S3T dynamics under homogeneous stochastic forcing admit, for all parameter values, a homogeneous equilibrium with zero mean flow and homogeneous eddy covariance:
	\begin{equation}
		\um^e=0\ \ ,\ \ C^e(\bx_a-\bx_b)=\frac{\e}{2}Q(\bx_a-\bx_b)\ .\label{eq:hom_eq}
	\end{equation}
The homogeneous equilibrium state~\eqref{eq:hom_eq} becomes unstable at certain parameter values and bifurcates to inhomogeneous equilibria, a
class of which are zonal jets.  
The stability of the homogeneous state~\eqref{eq:hom_eq} is addressed by linearizing~\eqref{eq:s3ts} around~\eqref{eq:hom_eq}.
Since~\eqref{eq:hom_eq} is homogeneous, the eigenfunctions
consist of a sinusoidal mean flow perturbation~$\d\um\,\ee^{\s t}$ and a perturbation covariance~$\d C\,\ee^{\s t}$ with a sinusoidal inhomogeneous part:
	\begin{equation}
		\d\um = \ee^{\i  n y}\ ,\ \ \d  C=\tilde{C}^{(h)}_{ n }(\bx_a-\bx_b)\,\ee^{\i  n (y_a+y_b)/2}\ ,\label{eq:eigfunc}
	\end{equation}
where $n$ is a real wavenumber that indicates the length-scale of the jets.  The corresponding eigenvalues $\s$ satisfy (see~\ifdraft Appendix~A\else\hyperref[app:dispersion]{Appendix~A}\fi):
	\begin{equation}
	  \s+1=f(\s|\d\um, C^e) = \e\,f(\s|\d\um, Q/2)\ ,\label{eq:dispersion}
	\end{equation}
where $f$ is the  vorticity flux induced by the distortion of the eddy equilibrium field $C^e$ by the mean flow $\d\um$; the expression for $f$ is given in~\eqref{eq:ff}.  This induced vorticity flux is referred to as the \emph{vorticity flux feedback on $\d\um$}.  For the ring forcing considered in this study, the fastest growing instability for $n<1$ has a real eigenvalue $\s$ and, therefore, the emergent jets are not translating in the $y$ direction.  The vorticity flux feedback at marginal stability
	\begin{equation}
		f_r\equiv \real\left[\bit f(\s=0|\d\um, Q/2 )\right]\ ,
	\end{equation}
that is positive in this case, has the tendency to reinforce the preexisting jet perturbation $\d\um$ and therefore destabilizes it.  With dissipation, the critical parameter $\e$ at which the homogeneous equilibrium
becomes unstable to a jet with wavenumber $n$ is $ \e_t(n)=1\big/f_r$ and for all values of $\beta$ there is a minimum energy input rate
	\begin{equation}
		\e_c\equiv\min_{ n }[\e_t(n)]\ ,
	\end{equation}
above which the homogeneous state is unstable and jet formation occurs.

It is instructive to identify which wave components (of the incoherent flow) contribute to the instability process.  For the forcing spectrum~\eqref{eq:Qhat} we may express the vorticity flux feedback at the stability boundary ($\s=0$) as
	\begin{equation}
  	\fr =\int_{0}^{\pi/2} \Fcal (\thet,n )\,\df\thet\ ,\label{eq:fr}
	\end{equation}
where $\Fcal(\thet,n)$ is the contribution to~$\fr$ from the wave components with wavevectors $\kv=(\pm\cos \thet,\pm\sin \thet)$
when the homogeneous equilibrium is perturbed by a jet perturbation with wavenumber~$n$.  Angle $\thet$
measures the inclination of the wave phase lines with respect to the $y$-axis.  The precise expression for $\Fcal(\thet,n)$ is
given in~\eqref{eq:Fcal}.  Positive values of $\Fcal$ indicate
that waves with phase lines inclined at angle $|\thet|$ produce up-gradient vorticity fluxes that are
destabilizing the jet perturbation $n$.  In general, destabilizing vorticity fluxes are produced by waves with phase lines closely aligned to
the $y$-axis (with small $|k_y|$) as shown in Fig.~\ref{fig:forcing}.

\begin{figure}[t]
\centerline{\includegraphics[width=19pc]{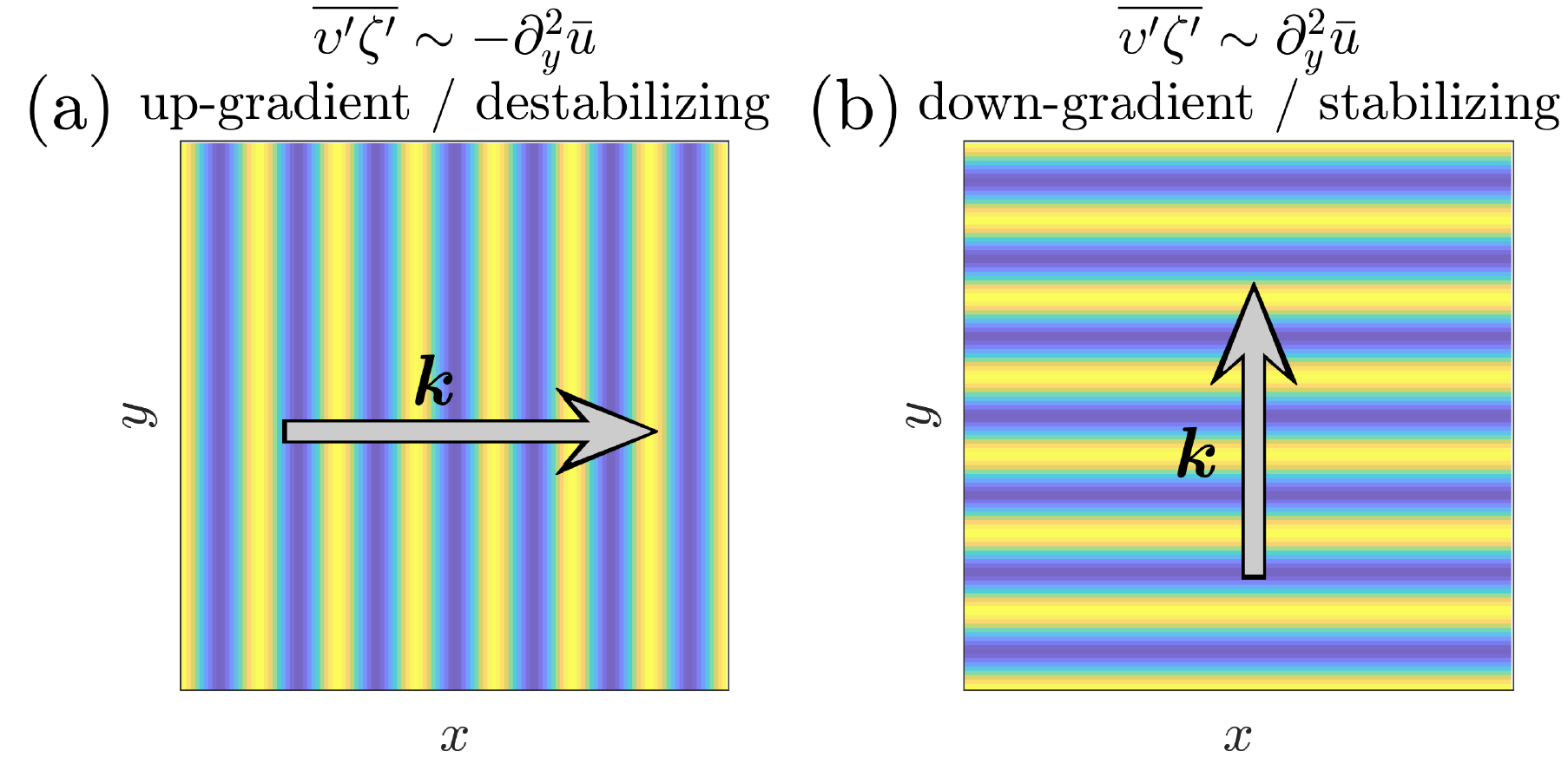}}
\vspace{-1em}
\caption{Waves with with small~$|k_y|$ (as in panel (a)) produce up-gradient vorticity fluxes that destabilize any mean flow perturbation superimposed on the homogeneous turbulent equilibrium; waves with large~$|k_y|$ (as in panel (b)) produce down-gradient vorticity fluxes that tend to diminish mean flow perturbations.} \label{fig:forcing}
\end{figure}

Figures~\ref{fig:flux_feedback}(a)~and~\ref{fig:flux_feedback}(b) show the contribution
$\Fcal(\thet,n)$ as a function of $\thet$ for the most unstable jet $n_c$ for the cases with $\beta=0.1$ and $\beta=100$. When $\beta\ll1$, $\Fcal(\thet,n)$ is positive for angles satisfying $4\sin^2\thet < 1+ n^2$.  This condition is derived
for $\beta=0$, but is also quite accurate for small $\beta$, as shown in Fig.~\ref{fig:flux_feedback}(a) \citep{Bakas-etal-2015}. The contribution from all angles is small (of order $\beta^2$), as the positive contribution at small angles is compensated by the negative contribution at larger angles.  For $\beta\gg 1$, only waves with phase lines almost parallel to the $y$ axis ($|k_y| \approx 0$) contribute significantly to the vorticity fluxes (see~Fig.~\ref{fig:flux_feedback}).  When integrated over all angles, the resulting vorticity flux feedback is positive and $O(\beta^{-2})$.  The wave--mean flow dynamics underlying these contributions at all values of $\beta$ can be understood by considering the evolution of wave groups in the sinusoidal flow and were studied in detail by~\cite{Bakas-Ioannou-2013-jas}.

\begin{figure}[t]
\centerline{\includegraphics[width=19pc]{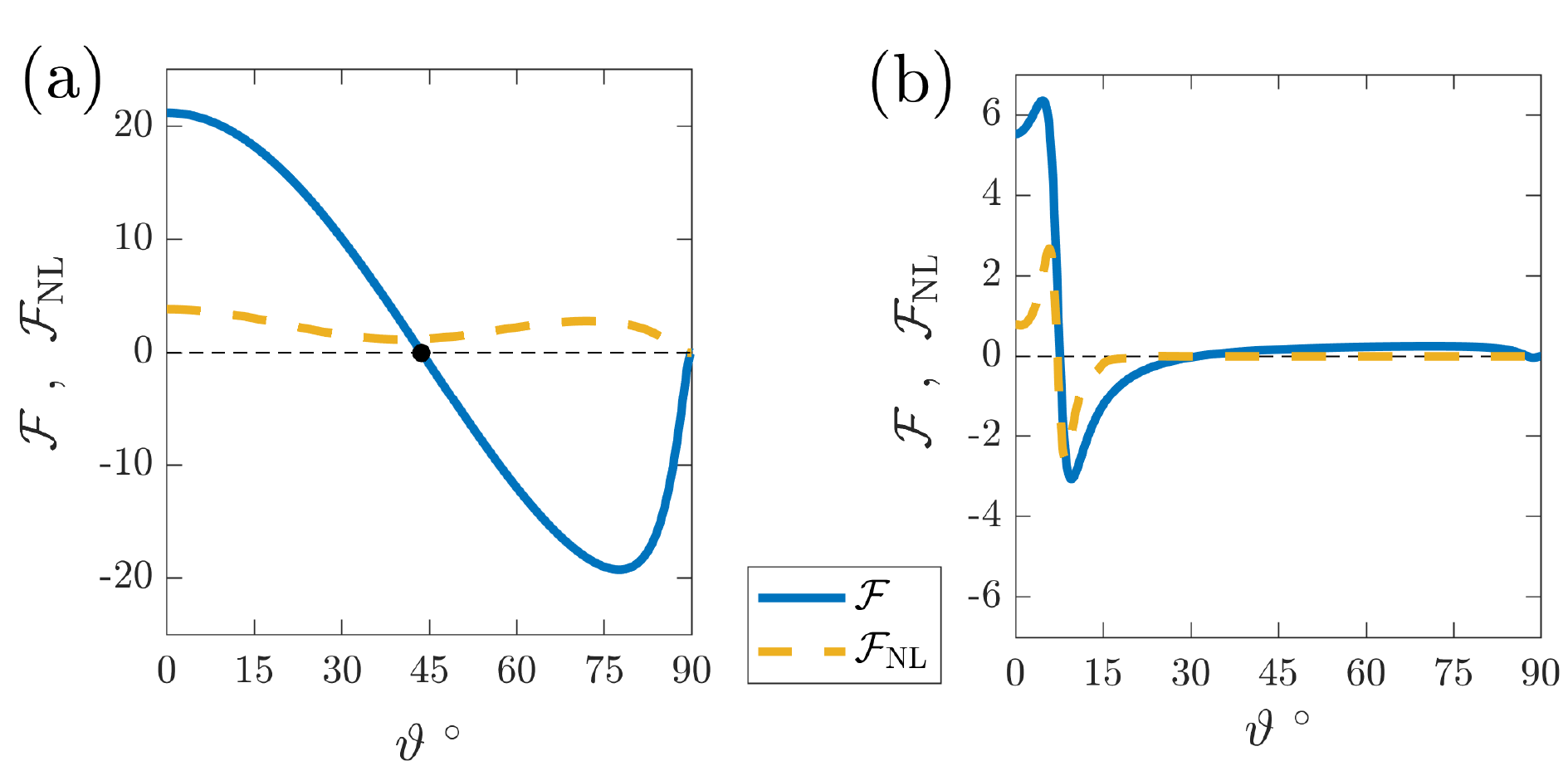}}
\vspace{-1em}
\caption{The contribution $\Fcal$ to the vorticity flux feedback $f_r$ for the most unstable jet eigenfunction from the waves with phase lines inclined at angle $\thet$ with respect to the meridional (solid curves).  Panel~(a) shows the case with $\beta=0.1$ while panel~(b) for $\beta=100$.  In panel~(a) the angle $\thet=\arcsin[\tfrac1{2}(1+ n ^2)^{1/2}]$ that separates the waves with positive (destabilizing) and negative (stabilizing) contribution to the vorticity flux feedback for $\beta=0$ is indicated with the filled circle.  Also, dashed curves show the contribution $\Fcal_{\rm NL}$ to the nonlinear Landau coefficient $c_3$ for the most unstable jet eigenfunction as a function of the wave angle $\vartheta$ (see section~\ref{sec:equil_landau}).} \label{fig:flux_feedback}
\end{figure}

\section{The Ginzburg--Landau (G--L) dynamics governing the nonlinear evolution of the flow-forming instability\label{sec:equil}}

In this section we discuss how the equilibration of the zonal jet instabilities is achieved for the case just above the critical threshold $\e_c$.  As it will be seen, the weak zonal jet equilibria are established  through the equilibration of the most unstable eigenfunction with wavenumber $n_c$ through a nonlinear feedback which modulates the eddy covariance in order to conserve energy and forms jet structures at the second harmonic $2n_c$.  It is through this energy conservation feedback along with the interaction with the $2n_c$ jet that equilibration is achieved.

To derive the asymptotic dynamics that govern the evolution of the jet amplitude we perform a multiple-scale perturbation analysis of the nonlinear dynamics near the marginal point.  Before proceeding with the multiple-scale analysis we present an intuitive argument that suggests the appropriate slow time and slow meridional spatial scales.

\subsection{The appropriate slow length scale and slow time scale}
For a stochastic excitation with energy input rate $\e=\e_c$, zonal jets with wavenumber $n=n_c$ are marginally stable.  If the energy input rate is slightly supercritical,
	\begin{equation}
		\e=\e_c(1+\mu^2)\ ,\label{eq:e_asympt}
	\end{equation}
with $\mu\ll 1$ a parameter that measures the supercriticality, then zonal jets with wavenumbers $|n - n_c|=O(\mu)$ are unstable and grow at a rate of $O(\mu^2)$.  To see this expand the eigenvalue relation~\eqref{eq:dispersion} near $\e_c$:
	\ifdraft
	\begin{align}
		\s=\mu^2\e_cf_r+\e_c  \left(\frac{\partial f}{\partial\s}\right)_c\s  + \frac{\e_c}{2}\left(\frac{\partial^2 f}{\partial  n ^2}\right)_c ( n - n_c )^2 + O\[\s^2,(n-n_c)^3\]\ ,\label{eq:dispersion2}
	\end{align}
	\else
	\begin{align}
		\s&=\mu^2\e_cf_r+\e_c  \left(\frac{\partial f}{\partial\s}\right)_c\s \nonumber\\
			&\quad + \frac{\e_c}{2}\left(\frac{\partial^2 f}{\partial  n ^2}\right)_c ( n - n_c )^2 + O\[\s^2,(n-n_c)^3\]\ ,\label{eq:dispersion2}
	\end{align}
	\fi
where the subscript $c$ denotes that the derivatives are evaluated at the threshold point $(\s,\mu,n)=(0,0,n_c)$.

Exactly at the minimum threshold, the function~$f$ has a maximum at $n=n_c$ ($(\partial f/\partial n)_c=0$ and
$(\partial^2f/\partial  n ^2)_c<0$) with value $\e_c f_r = 1$, which as seen from~\eqref{eq:dispersion} implies that $\s=0$.
Thus the approximate eigenvalue relation~\eqref{eq:dispersion2} predicts that the locus of points of marginal stability ($\s=0$) on the $\e$--$n$ plane lie on the parabola:
	\begin{equation}
		(n - n_c)^2 = \frac{2}{|f''_c| }\( \e\big/\e_c-1\) = \frac{2\mu^2}{|f''_c| }\ ,\label{eq:locus}
	\end{equation}
where $f''_c\equiv \left .  \partial^2 f / {\partial  n ^2} \right |_c$ and $\mu$ is the supercriticality parameter.

\begin{figure*}[ht]
\centerline{\includegraphics[width=39pc]{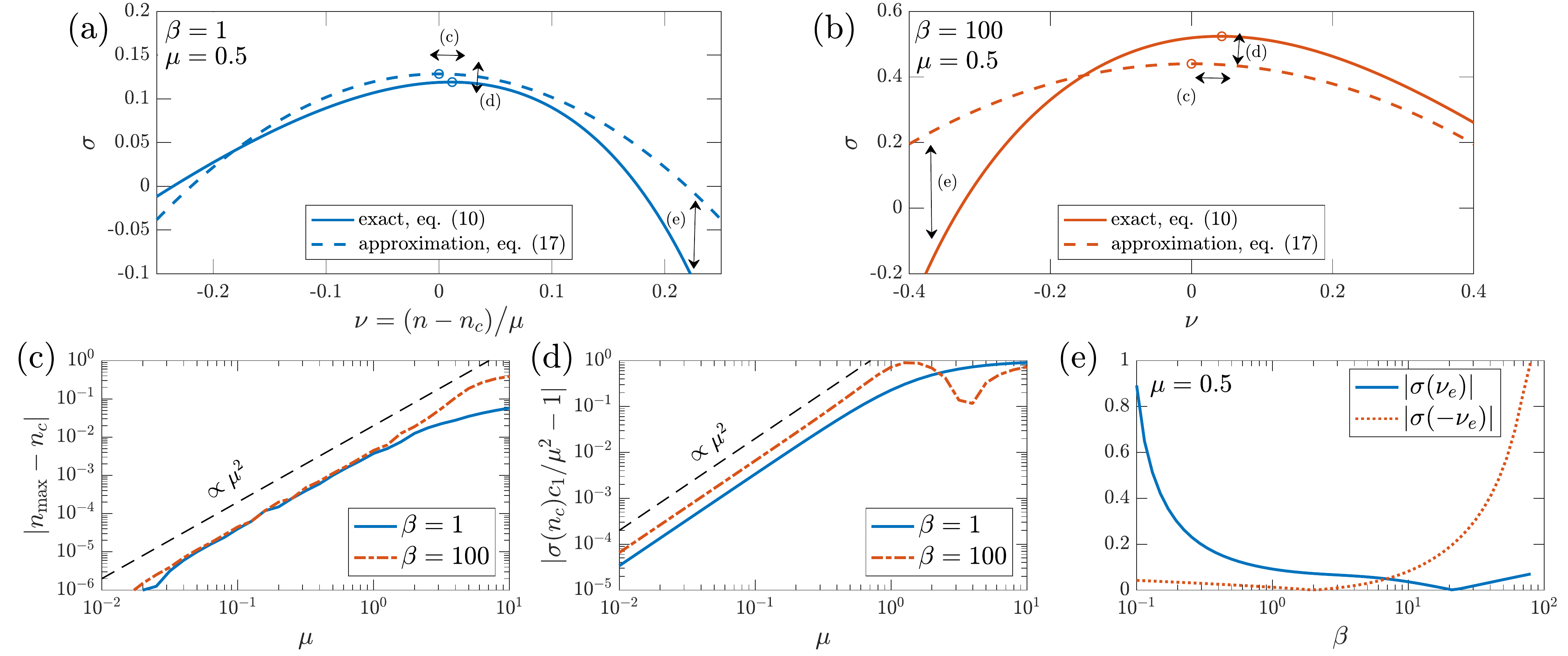}}
\vspace{-5mm}\caption{Validity of the approximate eigenvalue relation~\eqref{eq:stab}.  (a)~Comparison of the growth rates for jet perturbations with wavenumber $\nu$ as predicted by the exact eigenvalue relation~\eqref{eq:dispersion} (solid curve) and by the parabolic approximation~\eqref{eq:stab} (dashed curve) for supercriticality~$\mu=0.5$ and $\beta=1$.  Circles mark the maximum growth rate: for~\eqref{eq:dispersion} this is at wavenumber $n_{\mathrm{max}}$, while for~\eqref{eq:stab} at~$n_c$.  (b)~Same as panel (a) but for~$\beta=100$.  (c)~The difference between the exact wavenumber of maximum growth $n_{\mathrm{max}}$ and the approximate wavenumber of maximum growth~$n_c$ as a function of the supercriticality~$\mu$. (d)~The relative difference between the exact growth rate~$\s$ for a jet at wavenumber~$n_c$ and the approximate growth rate~$\mu^2/c_1$ as a function of supercriticality~$\mu$. (e)~The exact growth rate of jet perturbations with wavenumbers $\nu_e$ and $-\nu_e$ as a function of~$\beta$ for supercriticality~$\mu=0.5$.  The parabolic approximation predicts zero growth for these marginal wavenumbers.}\label{fig:disp}
\end{figure*}

Using~\eqref{eq:dispersion2} we can estimate the growth rate $\s$ at supercriticality $\mu$.  We find that jets with wavenumber $n = n_c + \mu\nu$ grow approximately at rate:
	\begin{equation}
		\s = \mu^2 (1 -c_2 \nu^2)\big/c_1\ ,\label{eq:stab}
	\end{equation}
with
	\begin{equation}
		c_1 \equiv 1-\e_c\left(\frac{\partial f}{\partial\s}\right)_{c}\quad\text{and}\quad c_2\equiv\frac{\e_c}{2} |f''_c|\ .\label{def_c1c2}
	\end{equation}
The analytic expressions for $c_1$ and $c_2$ are given in~\eqref{eq:c1} and~\eqref{eq:c2}.  Coefficient $c_1$ is positive for stochastic excitations with spectrum~\eqref{eq:Qhat}.  From~\eqref{eq:stab}, we deduce that for any~$\mu$ only jets with
	\begin{equation}
		|\nu|<\nu_e\defn 1\big/ \sqrt{c_2}\ ,\label{eq:nucrit}
	\end{equation}
can become unstable.

Equations~\eqref{eq:locus} and~\eqref{eq:stab} establish the initial assertion: for $\mu\ll1$ zonal jets with wavenumbers $|n-n_c|=O(\mu)$ grow at a rate $\s=O(\mu^2)$.  

The validity of the approximate eigenvalue relation~\eqref{eq:stab} as a function of supercriticality $\mu$ is shown in panels~(a) and~(b) of Fig.~\ref{fig:disp}.  By comparing the exact growth rates as given by~\eqref{eq:dispersion} and the growth rates obtained from the approximation~\eqref{eq:stab}, we see that the approximate eigenvalue dispersion may not be as accurate in three ways: predicting the maximum growth rate, predicting the wavenumber at which maximum growth occurs, and predicting the asymmetry of the exact growth rates about the maximal wavenumber.
These three differences are indicated by the arrows in panels~(a) and~(b) of Fig.~\ref{fig:disp} and are quantified in panels~(c) through~(e).
Panel~(c) compares the exact wavenumber of maximum growth $n_{\rm max}$ to the critical wavenumber $n_c$ assumed by approximation~\eqref{eq:stab}.
We see that $n_{\rm max}$ is very close to $n_c$ up to $\mu\approx 1$ with the error growing as $\mu^2$.  This is in agreement with the error
in~\eqref{eq:dispersion2} being of $O(\s^2)$.  In addition, the exact growth rate $\s(n_c)$ is very close to $\mu^2/c_1$, as shown in Fig.~\ref{fig:disp}(d) for $\mu$ up to $O(1)$; the growth rate being overestimated by~\eqref{eq:stab} for higher values.  Finally, the parabolic approximation~\eqref{eq:stab} to the growth rates predicts that the wavenumbers $\pm\nu_e$ are marginally stable ($\s(\pm \nu_e) =0$).  Figure~\ref{fig:disp}(e) shows the exact growth rates at $\pm\nu_e$ at $\mu=0.5$; these are far from zero for both low and high values of $\beta$.  The parabolic approximation works best for intermediate range $\beta$ values, i.e.,~for $\beta=O(1)$.  To summarize, the approximated maximum growth rate $\mu^2/c_1$ as well as the the critical wavenumber $n_c$ that achieves this maximum growth are both good approximations for supercriticalities up to $\mu=O(1)$; the parabolic dependence of growth rate for wavenumbers away from $n_c$ is a good approximation at $\mu>0.1$ only for intermediate values of $\beta$.  As it will be seen, this has implications on the validity of the weakly nonlinear dynamics derived next.

\subsection{\mbox{G--L} dynamics for weakly supercritical zonal jets}

Since the excess energy available for flow formation is of order $\mu^2\e_c$, we expect intuitively the mean flow
amplitude to be of order $\mu$. Therefore, to obtain the dynamics that govern weakly supercritical zonal flows, we expand the mean flow $\um$ and the covariance $C$ of the S3T equations~\eqref{eq:s3ts} as:
	\begin{subequations}
	\label{eq:pert}
	\ifdraft
	\begin{align}
  	\um & = \mu \um_1(y,Y,T)+\mu^2\um_2(y,Y,T)+O(\mu^3)\ ,\label{eq:Uexp}\\
    C & = C^e(\bx_a-\bx_b)+\mu C_1(\bx_a,\bx_b,Y_a,Y_b,T) +\mu^2 C_2(\bx_a,\bx_b,Y_a,Y_b,T)+O(\mu^3)\ , \label{eq:Cexp}
	\end{align}
	\else
	\begin{align}
  	\um & = \mu \um_1(y,Y,T)+\mu^2\,\um_2(y,Y,T)+O(\mu^3)\ ,\label{eq:Uexp}\\
  	C & = C^e(\bx_a-\bx_b)+\mu\, C_1(\bx_a,\bx_b,Y_a,Y_b,T) \nonumber\\
  		&\qquad+\mu^2\, C_2(\bx_a,\bx_b,Y_a,Y_b,T)+O(\mu^3)\ .  \label{eq:Cexp}
	\end{align}
	\fi
	\end{subequations}
Guided by~\eqref{eq:locus} and~\eqref{eq:stab}, we have assumed that the zonal jet and its associated covariance evolve from the marginal values at the slow time scale $T\equiv\mu^2 t$, while being modulated at the long meridional scale $Y\equiv \mu y$.

Details of the perturbation analysis are given in the \ifdraft Appendix~B\else\hyperref[app:GL]{Appendix~B}\fi; here we present the backbone.
We introduce~\eqref{eq:pert} in~\eqref{eq:s3ts} and gather terms with the same power of $\mu$.  At leading order $\mu^0$, we recover the homogeneous equilibrium~\eqref{eq:hom_eq}.  At order $\mu^1$, the emergent zonal jet and the covariance are the modulated S3T eigenfunction:
	\begin{subequations}
	\ifdraft
	\begin{align}
  	\um_1&=A(Y,T)\,\ee^{\i  n_c y}+\mbox{c.c.}\ , \\
  	C_1&=\left[ \bit A(Y_a,T)G^+(0|\bx_a-\bx_b) \right.  -A(Y_b,T)G^-(0|\bx_a-\bx_b)\left.\bit\right]\ee^{\i  n_c (y_a+y_b)/2}+\mbox{c.c.}\ ,\label{eq:ordermu}
	\end{align}
\else
	\begin{align}
  	\um_1&=A(Y,T)\,\ee^{\i n_c y}+\mbox{c.c.}\ , \\
  	C_1&=\left[ \bit A(Y_a,T)\,G_c^+(0|\bx_a-\bx_b) \right.  \label{eq:ordermu}\\
  		&\qquad \left.-A(Y_b,T)\,G_c^-(0|\bx_a-\bx_b)\bit\right]\ee^{\i n_c (y_a+y_b)/2}+\mbox{c.c.}\ ,\nonumber
	\end{align}
	\fi
	\end{subequations}
with $G_c^{\pm}$ defined in~\eqref{eq:G} and evaluated at $n=n_c$.

Having determined~$C_1$ we proceed to determine the order $\mu^2$ correction of the covariance, $C_2$.  This step of the calculation is facilitated if we disregard the dependence on the slow spatial scale~$Y$ in the amplitude~$A$, as well as that in~$C_1$ and $C_2$.  \cite{Parker-Krommes-2014-generation} showed that the nonlinear term of the asymptotic dynamics responsible for the equilibration of the amplitude $A$ can be obtained using this simplification, while the contribution to the asymptotic dynamics from the
slow varying latitude $Y$ is the addition of a diffusion term with the diffusion coefficient $c_2$ in~\eqref{def_c1c2}.
At order~$\mu^2$ a zonal jet with wavenumber~$2n_c$ emerges:
	\begin{subequations}
	\begin{equation}
		\um_2=\alpha_2\,A(T)^2\,\ee^{2\i n_c y}+\mbox{c.c.}\ ,\label{eq:2nc}
	\end{equation}
where $\alpha_2$ is given in~\eqref{eq:a1} and for the forcing considered is negative ($\alpha_2<0$).  The associated covariance at order $\mu^2$,
\ifdraft
	\begin{align}
	  C_2&= C^e (\bx_a-\bx_b)+C_{20}(\bx_a-\bx_b)+C_{22}(\bx_a-\bx_b)\,\ee^{2\i n_c (y_a+y_b)/2}+\mbox{c.c.}\ .
	\label{eq:C2}
	\end{align}
\else
	\begin{align}
	  C_2&= C^e (\bx_a-\bx_b)+C_{20}(\bx_a-\bx_b,T) \nonumber\\
	  &\qquad+C_{22}(\bx_a-\bx_b,T)\,\ee^{2\i n_c (y_a+y_b)/2}+\mbox{c.c.}\ ,
	\label{eq:C2}
	\end{align}\fi
	\end{subequations}
consists of the homogeneous part, $C^e + C_{20}$, and also an inhomogeneous contribution at wavenumber $2n_c$.  (Note that, as implied by~\eqref{eq:e_asympt}, the forcing covariance $Q$ appears both at order~$\mu^0$ and at order $\mu^2$.)

The homogeneous covariance contribution, $C^e+C_{20}$, is required at order $\mu^2$ so that the energy
conservation~\eqref{eq:tot_en} is satisfied.  To show this note that as the instability develops on a slow time scale, the total
energy density has already assumed (over an order one time scale) its steady state value  $\varepsilon/2$ (see~\eqref{eq:tot_en})
and therefore, the mean flow energy growth \emph{must} be accompanied by a decrease in the eddy energy.  This decrease is facilitated
by a concomitant change of the eddy covariance at order $\mu^2$.  Specifically,
by introducing perturbation expansion~\eqref{eq:pert} in~\eqref{eq:tot_en} at steady state, we obtain at leading order, $\mu^0$, the trivial balance:
	\begin{equation}
	  -\dashint\df^2\bx\;\frac1{4}\left[(\Del_a^{-1}+\Del_b^{-1})
	  C^e\right]_{a=b} =  \frac{\e_c}{2}\ .\label{eq_E^0}
	\end{equation}
At order $\mu^1$ the eddy covariance does not contribute to the energy since $C_1$ is harmonic in $y$ and integrates to zero:
	\begin{equation}
	  \dashint\df^2\bx \;\frac1{4}\left[(\Del_a^{-1}+\Del_b^{-1})C_1\right]_{a=b} =0\ .
	\end{equation}
At order $\mu^2$ we use \textit{(i)}~\eqref{eq_E^0} and \textit{(ii)}~that the inhomogeneous component
$C_{22}\,\ee^{2\i n_c (y_a+y_b)/2}$ is harmonic and integrates to zero, to obtain:
	\begin{equation}
		\dashint\df^2\bx\;\frac1{4}\left[(\Del_a^{-1}+\Del_b^{-1}) C_{20}  \right]_{a=b} = \dashint\df^2\bx\;\frac1{2}\um_1^2\ .\label{eq:conC20}
	\end{equation}
Thus the homogeneous deviation from the equilibrium covariance must produce a perturbation energy \emph{defect} to 
counter balance the energy growth of the mean flow.  We refer to $C_{20}$ as the \emph{eddy energy correction term}. 
However, we note that the correction to the homogeneous part of the covariance does not only change the mean eddy energy but also other eddy characteristics, such as the mean eddy anisotropy, that also might play a role in the equilibration process.

At order $\mu^3$  secular terms appear which, if suppressed, yield an asymptotic perturbation expansion  up to
time $O(1/\mu^2)$.  Suppression of these secular terms requires that the amplitude $A$ of the most unstable jet with wavenumber~$n_c$ satisfies:
	\begin{equation}
		c_1\,\partial_T A=\,A-c_3\,A|A|^2\ .\label{eq:GL0}
	\end{equation}
If we now allow the amplitude to also evolve with the slow scale, $Y$, and add the diffusion term $c_2\partial_{Y}^2 A$ on the right-hand-side of~\eqref{eq:GL0}, we obtain the real Ginzburg--Landau (\mbox{G--L}) equation:
	\begin{equation}
c_1\,\partial_T A=\,A+c_2\,\partial_{Y}^2A-c_3\,A|A|^2\ .\label{eq:GL}
\end{equation}
For forcing with spectrum~\eqref{eq:Qhat} all three coefficients $c_1$, $c_2$, and~$c_3$ are real and positive.  The coefficients $c_1$ and $c_2$, are the coefficients in the  Taylor  expansion~\eqref{eq:dispersion2} and are given in~\eqref{def_c1c2}.

The \mbox{G--L} equation~\eqref{eq:GL} has a steady solution $A=0$.  This solution is linearly unstable to modal perturbations
$\ee^{\i \nu Y + \s T}$, with growth rate $\mu^2(1 - \nu^2 c_2)/c_1$; the most unstable mode occurs at $\nu=0$.  This is the flow-forming
SSD instability of the homogeneous equilibrium state in the \mbox{G--L} framework (cf.~\eqref{eq:stab}).  The \mbox{G--L}
equation has also the nonlinear harmonic equilibria
	\begin{equation}
  	A(Y)=R_0(\nu)\,\ee^{\i(\nu Y+\varphi)}~~{\rm with}~~R_0(\nu)=\sqrt{(1-\nu^2c_2)\big/c_3}\ ,\label{eq:eq_amp}
	\end{equation}
and $\varphi$ an undetermined phase that reflects the translational invariance of the system in $y$.  These equilibria are the possible finite-amplitude jets that emerge at low supercriticality.  However, as will be shown in the next section, some of these equilibria are susceptible to a secondary SSD instability and evolve through jet merging or jet branching to the subset of the stable attracting states.

The \mbox{G--L} equation obeys potential dynamics and thus the system always ends up in a stationary state which is a local minimum of the potential \citep{Cross-Greenside-2009}.  The $\nu=0$ jet is the state that corresponds to the global minimum of the potential and it has amplitude
\begin{equation}
 R_0(0)=1\big/\sqrt{c_3}\ .\label{eq:eq_amp0}
\end{equation}


\section{Comparison of the predictions of \mbox{G--L} dynamics with S3T dynamics for the equilibrated jets\label{sec:compare}}

In this section we test the validity of the weakly nonlinear \mbox{G--L} dynamics by comparing its predictions for the amplitude of the equilibrated jets with fully nonlinear S3T dynamics.  We consider the S3T dynamical system~\eqref{eq:s3ts} in a doubly periodic domain $2\pi\Ld\times 2\pi\Ld$ with a $128^2$ grid-resolution and $\Ld=1$, as well as the \mbox{G--L} dynamics with periodic boundary conditions for the amplitude of the jet, $A$, on the same domain.  We approximate the delta function in the ring forcing~\eqref{eq:Qhat} with
	\begin{equation}
		\delta(\kd-\kfd)\mapsto \frac{\ee^{-(\kd-\kfd)^2/(2\,\dfd^2)}}{\sqrt{2\pi} \dfd}\ ,\label{eq:Qhat_disc}
	\end{equation}
with $\kxd\Ld$, $\kyd\Ld$ assuming integer values. (The asterisks denote dimensional values, as in, e.g.,~\eqref{eq:NLbarotropic}.) Forcing~\eqref{eq:Qhat_disc} injects energy in a narrow ring in wavenumber space with radius $\kfd\Ld=10$ and width $\dfd\Ld=1.5$.  We note that even
though~\eqref{eq:Qhat_disc} is a good approximation of the delta-ring forcing~\eqref{eq:Qhat}), small quantitative differences are to be expected.  For example, the critical energy input rates for jet emergence obtained from the discrete finite ring excitation differ by as much as 4\% from the corresponding values obtained from the delta-ring forcing~\eqref{eq:Qhat}.  Since the equilibrated jet amplitudes are of order $\mu\ll1$, we use the exact values for the critical energy input rates obtained for the discrete finite ring excitation.

We also consider $\rd=0.1$ and vary $\betad$ as well as the energy input rate $\ed$ that is the bifurcation parameter.  The eigenvalue relation for the flow-forming instability can be readily obtained by substituting the integrals in~\eqref{eq:dispersion} with sums over the allowed wavenumbers.  However, the comparison with the predictions of the \mbox{G--L} dynamics with periodic boundary conditions in the meridional is more tricky.  Due to the periodic boundary conditions in the jet amplitude, a harmonic mean flow
$A(Y,T)=\ee^{\i \nu Y}$ with wavenumber $\nu$ corresponds to a mean flow within our domain \emph{only} if its dimensional wavenumber
$n_*=(\mu \nu + n_c)\kfd$ is an integer.  Therefore, we carefully pick $\betad$ so that the marginal wavenumber $n_c \kfd$ always assumes an integer value; for $\kfd=10/L_*$ this leaves us with nine possible values for $\betad$ covering the range $3\times10^{-1}<\betad<2\times10^{3}$.  The lowest and highest marginal $\betad$-values yield marginal jets at the lowest and highest
allowed wavenumber possible within our domain; $1/L_*$ and $\kfd -1/\Ld$ respectively.  We excluded these values for $\betad$, since
they do not allow us to study the finite amplitude stability of side-band jets (i.e.,~jets at larger or smaller scale compared to the scale of
$\ncd$).  Therefore, in our comparisons we use only the remaining seven values of $\betad$, which are shown in Table~\ref{table:b_values}.

\begin{table}
\footnotesize\caption{Exact values of non-dimensional planetary vorticity gradient $\beta \equiv \betad/(\kfd\rd)$ used in the S3T simulations of section~\ref{sec:compare} and their corresponding values of the dimensional critical wavenumber $\ncd$.}
\label{table1}\centering
\begin{tabular}{c c c}
   Notation   &   $\beta$  & $\ncd$\\
	\hline
$\beta_1$ & 1.1915 & 8\\
$\beta_3$ & 3.0235 & 7\\
$\beta_6$ & 6.2761 & 6\\
$\beta_{12}$ & 12.136 & 5\\
$\beta_{24}$ & 24.576 & 4\\
$\beta_{58}$ & 58.137 & 3\\
$\beta_{192}$ & 192.62 & 2\\
\hline
\end{tabular}\label{table:b_values}
\end{table}

\begin{figure*}
\centerline{\includegraphics[width=22pc]{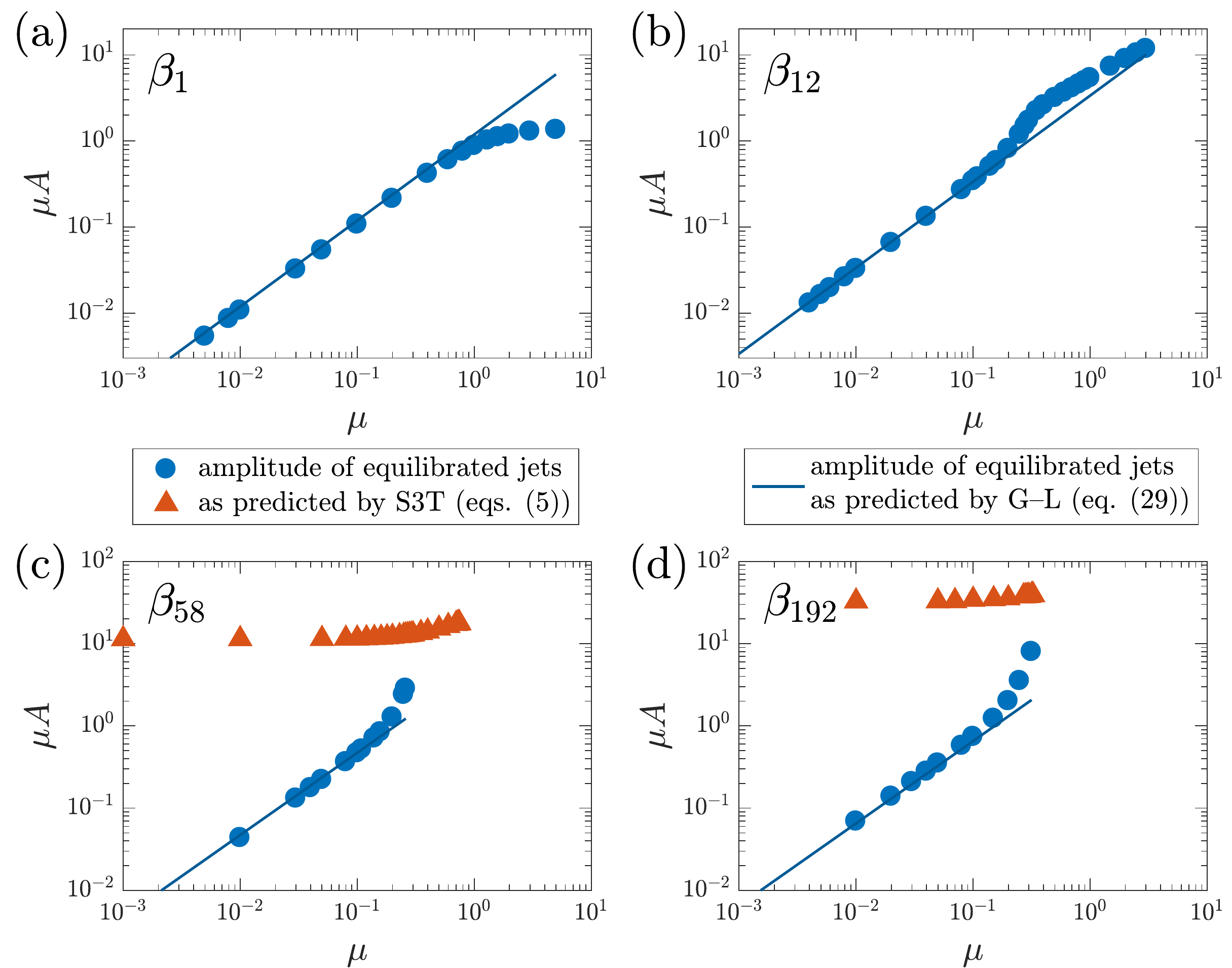}}
\vspace{-1em}
\caption{The amplitude $\mu A$ of the equilibrated most unstable jet with wavenumber $n_c$ as a function of supercriticality $\mu$ for four values of $\beta$.  The \mbox{G--L} branch is shown with circles $\bigcirc$; the upper branch (which appears for $\beta\gtrapprox 20$) is shown with triangles $\triangle$.  Solid lines show the jet amplitude as predicted by the \mbox{G--L} (cf.~\eqref{eq:eq_amp0}).}
\label{fig:nc_ampl1and2}
\centerline{\includegraphics[width=39pc]{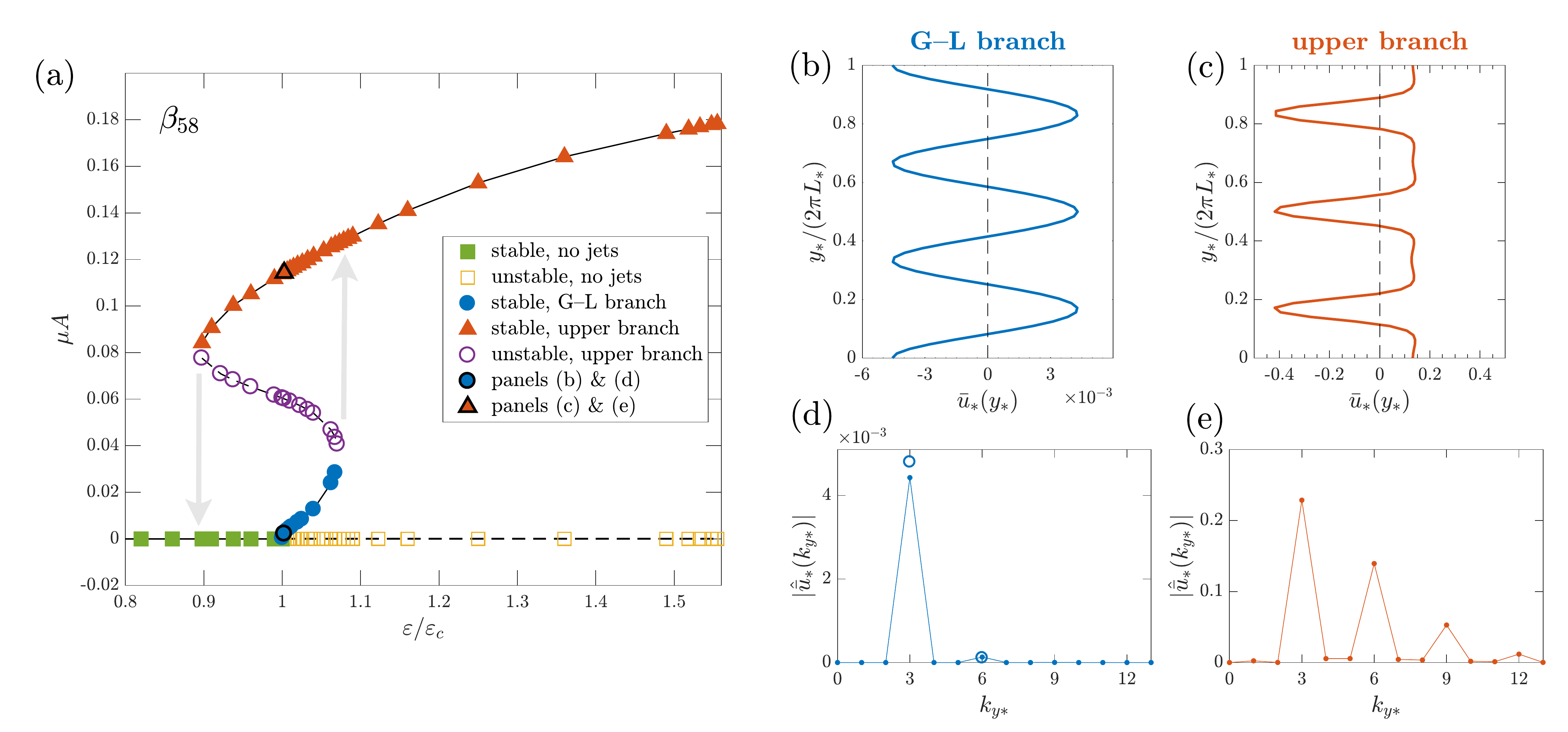}}
\vspace{-1em}
\caption{The bifurcation diagram for $\beta_{58}$ (case shown in panel (c) of Fig.~\ref{fig:nc_ampl1and2}).  (a) The amplitude $\mu A$ of the
equilibrated most unstable jet with wavenumber $n_c$ as a function of the energy input rate.  Squares denote the homogeneous equilibrium,
circles the lower branch predicted by the \mbox{G--L} dynamics and the triangles the upper branch of equilibria.  Open symbols denote
unstable jet equilibria with respect to S3T dynamics; filled symbols denote stable jet equilibria.  Multiple stable equilibria exist
for $0.89\le \e/\e_c\le 1.068$.  A comparison of the jet equilibrium structure and the jet spectra for $\e=1.0025$ (which corresponds to $\mu=0.05$) is shown in panels (b)-(d).  Panels (b), (d) show the lower \mbox{G--L} branch jet; Panels (c), (e) show the upper branch jet.  In panel (d) the amplitude prediction for $n_c$ by \eqref{eq:eq_amp0} and for $2n_c$ by \eqref{eq:2nc} is also shown (open circles).}
\label{fig:bifurc_b58_jetspectra}
\end{figure*}

We calculate the finite amplitude equilibrated jets from the nonlinear S3T dynamical system~\eqref{eq:s3ts} using a Newton's method with the initial guess provided by~\eqref{eq:eq_amp0}.\footnote{For details regarding Newton's algorithm for system~\eqref{eq:s3ts} the reader is referred to the Appendix~I in the thesis of~\citet{Constantinou-2015-phd}.} All jet equilibria we compute in this section are hydrodynamically stable. At small supercriticalities the jet amplitude is small and the linear operator is dominated by dissipation. Thus, all instabilities we discus here are ``SSD instabilities'' (see the discussion in \S3 of section~\ref{sec:intro}).

\subsection{Equilibration of the most unstable jet, $n_c$}

Consider first the most unstable jet perturbation with wavenumber $n_c$.  Figure~\ref{fig:nc_ampl1and2} shows the
Fourier amplitude of the equilibrated jet dominated by wavenumber $n_c$ for four values of~$\beta$.  We see that for $\beta\lessapprox 12$,
the amplitude is given, to a very good approximation, by~\eqref{eq:eq_amp0} for supercriticality up to $\mu\approx 0.2$ (see panels (a)-(b)).
For larger supercriticality, the amplitude of the equilibrated jet is not well captured by~\eqref{eq:eq_amp0}; the jet amplitude is
overestimated for $\beta\lessapprox 12$ while it is underestimated for $\beta\gtrapprox 12$.  We note here, that S3T equilibria with
dominant wavenumber $n_c$ (as predicted by the \mbox{G--L} dynamics) exist at even larger supercriticalities but these were found to be S3T
unstable.

Surprisingly, for $\beta\gtrapprox20$ there exist multiple equilibria for the same supercriticality~$\mu$ (see panels (c)-(d) of Fig.~\ref{fig:nc_ampl1and2}).  Specifically, there exists a branch of stable equilibria apart from the jets connected to the homogeneous equilibrium (cf.~triangles in Fig.~\ref{fig:nc_ampl1and2}(c)-(d) versus the circles).  For $\mu \gtrapprox 0.2$, the lower branch equilibria, predicted by the \mbox{G--L} dynamics, do not exist; an infinitesimal harmonic jet perturbation with wavenumber $n_c$ ends
up in the upper branch.  Equally interesting is the fact that  the upper branch extends to subcritical values of the energy input rate with respect to the flow-forming instability of the homogeneous state, i.e.~for $\e<\e_c$.  This is shown in Fig.~\ref{fig:bifurc_b58_jetspectra}(a) for $\beta_{58}$ and similar subcritical jet equilibria were found for $\beta_{24}$ and $\beta_{192}$ (not shown).  Thus, apart from the linear instability forming jets that has been extensively studied in the literature, there is a nonlinear instability for jet formation the details of which will be discussed in section~\ref{sec:equil_landau}\ref{sec:upper_branch}.  Since both the upper and the lower branch exist for a limited range of energy input rates, there is a hysterisis loop shown in Fig.~\ref{fig:bifurc_b58_jetspectra}(a), with the dynamics landing on the upper or the lower branch of jet equilibria as $\varepsilon$ is varied.  The two stable branches are connected with a branch of unstable equilibria (open circles) that were also found using Newton's method.

The jets on the lower and the upper stable branch are qualitatively different.  Panels (b)-(e) of Fig.~\ref{fig:bifurc_b58_jetspectra} compare the jet structure and spectra of two such equilibria in the case of $\beta_{58}$ and $\mu=0.05$.  While the lower branch jet consists mainly of $n_c$ and its double harmonic, $2n_c$, with a much weaker Fourier amplitude in qualitative agreement with the \mbox{G--L} prediction of $\overline{u}_2 \approx O(\mu^2A^2)$ (c.f.~\eqref{eq:2nc}), the upper branch jet is stronger by two orders of magnitude, it contains more harmonics and the Fourier amplitude of the double harmonic $2n_c$ is about half the amplitude of the leading harmonic $n_c$.  As will be elaborated in section~\ref{sec:equil_landau}\ref{sec:upper_branch}, it is the interaction of the two jets with wavenumbers $n_c$ and $2n_c$ that supports the upper branch equilibria.


\subsection{Equilibration of the side-band jets, $n_c\pm 1/(\kfd\Ld)$}

We now consider the jet equilibria that emerge from the equilibration of jet perturbations with wavenumbers close to $n_c$.  While for an infinite domain there is a dense set of unstable jet perturbations with wavenumbers $\mu\nu$ close to $n_c$ (cf.~\eqref{eq:nucrit}), for the doubly periodic box the first side band jet instabilities have dimensional wavenumbers $n_*^\pm=n_c\kfd\pm 1/\Ld$, or $\nu^\pm=\pm 1/(\kfd\Ld\mu)$.  Introducing $\nu^\pm$ in~\eqref{eq:nucrit}, we obtain that the parabolic approximation predicts that the homogeneous equilibrium becomes unstable to jet perturbations with
wavenumber $\nu^\pm$ when $\mu_{\mathrm{GL}}>\sqrt{c_2}/(\kfd\Ld)$.  However,
as shown in Fig.~\ref{fig:disp}(d), the parabolic approximation is not accurate especially at low and large values of $\beta$.  For example for $\beta_1$, $\mu_{\mathrm{GL}}=0.4293$, while the exact dispersion relation predicts that jets with $\nu_+$ and $\nu_-$ are rendered neutral at $\mu_{\mathrm{ex}^+}=0.2140$ and $\mu_{\mathrm{ex}^-}=0.7953$ respectively.  We therefore expect significant deviations from~\eqref{eq:eq_amp} for the amplitude of the equilibrated jets.

Figure~\ref{fig:sideband_eq} shows the equilibrated amplitude of the side band jet perturbations with $\nu^\pm$ as a function of supercriticality for four values of $\beta$.  While the functional dependence of the equilibrated amplitude on $\mu$ is qualitatively captured by~\eqref{eq:eq_amp} (dashed lines), there are significant quantitative differences especially for $\beta_1$ and $\beta_{192}$.  Since these quantitative differences are due to the failure of the parabolic approximation, a way to rectify them is to use an equivalent
	\begin{equation}
		c_2^{\mathrm{ex}\pm} \defn (\kfd\Ld\mu_{\mathrm{ex}^\pm})^2\ ,
	\end{equation}
based on the supercriticality $\mu_{\mathrm{ex}}^\pm$ obtained from the exact dispersion relation~\eqref{eq:dispersion}.  The solid curves show the predicted amplitude using $c_2^{\mathrm{ex}\pm}$.  We observe that for all values of $\beta$ the amplitude of the jets close to the bifurcation point is accurately predicted and for the intermediate value of $\beta_6$ for which the
exact dispersion is the closest to the parabolic profile, the agreement holds away from the bifurcation point as well. Finally, note that for large
$\beta$ shown in Fig.~\ref{fig:sideband_eq}(c) the additional upper branch of equilibria is found and has the same characteristics as the upper
branch of $n_c$ equilibria. That is, the equilibrated jets have a larger amplitude and the Fourier amplitude of the double harmonic (in this case
it is the $2(n_c+1/\mu\kfd\Ld)$ harmonic) is much larger compared to the \mbox{G-L} branch.

Finally, we stress that the results in this section regarding the existence of the upper branch equilibria as well as the accuracy  of the G--L dynamics for the lower branch equilibria are not quirks of the particular isotropic forcing structure in~\eqref{eq:Qhat} but rather similar qualitative behavior is found for forcing with anisotropic spectrum. Discussion regarding the effects of the structure of the forcing is found in \ifdraft Appendix~C\else\hyperref[app:NIF]{Appendix~C}\fi.

\begin{figure*}
\centerline{\includegraphics[width=0.9\textwidth]{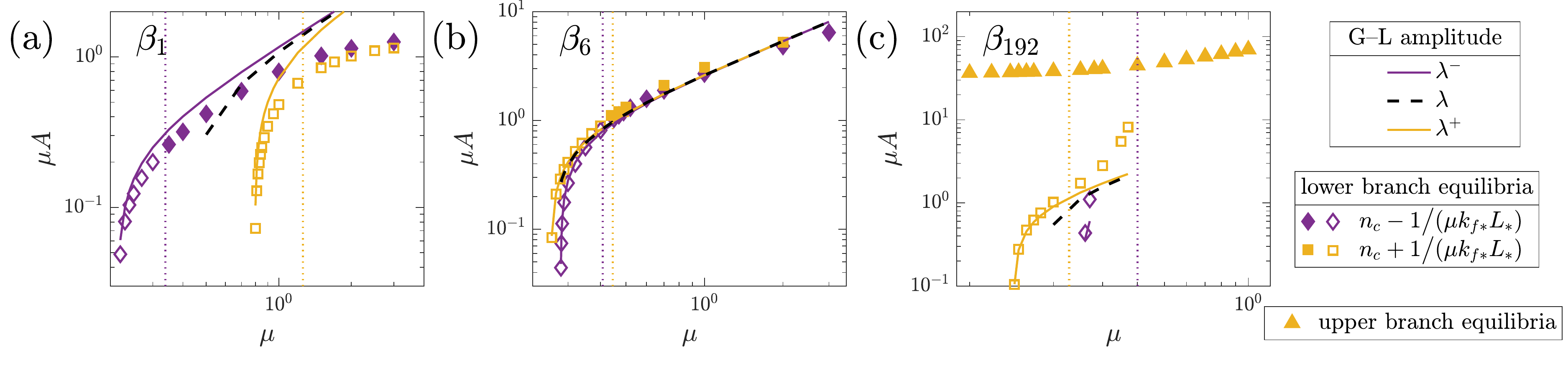}}
\vspace{-1em}
\caption{The amplitude $\mu A$ of the equilibrated unstable jets with wavenumbers $n_c-1/(\mu \kfd\Ld)$ (diamonds) and $n_c+1/(\mu \kfd\Ld)$ (squares) as a function of supercriticality $\mu$ for four values of $\beta$.  The dashed lines show the amplitude predicted by the \mbox{G--L} dynamics (cf.~\eqref{eq:eq_amp0}), while the solid lines show the amplitude predicted by the \mbox{G--L} dynamics with $c_2^{\mathrm{ex}\pm}$ as described in the text.  Stable (unstable) equilibria are denoted with filled
(empty) symbols and the vertical dotted lines show the stability boundary \eqref{eq:stab_bound} obtained from the \mbox{G--L} dynamics (see section~\ref{sec:Eckhaus}).}
\label{fig:sideband_eq}
\end{figure*}

\begin{figure}
\centerline{\includegraphics[width=19pc]{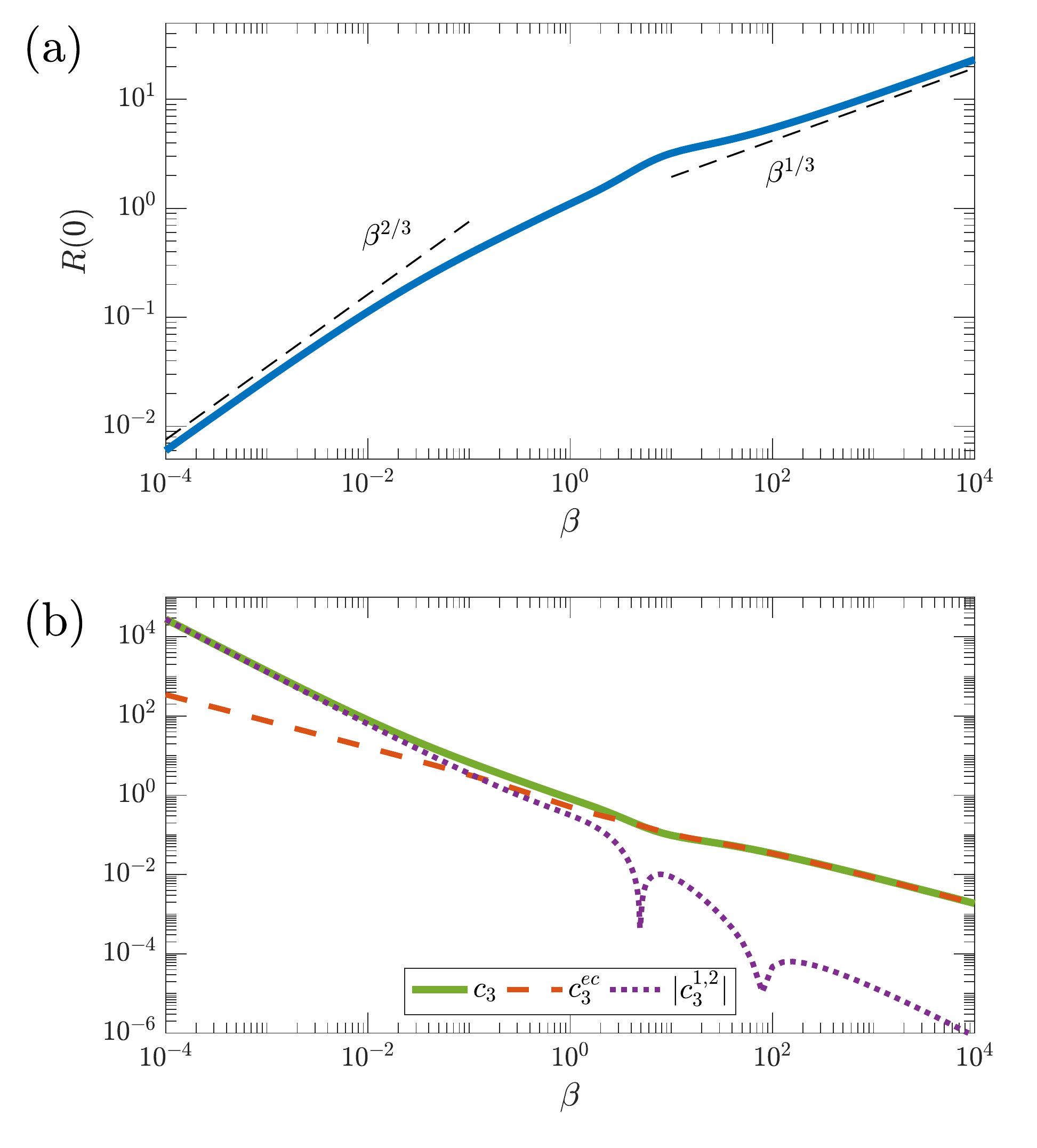}}
\vspace{-1em}
\caption{(a) The amplitude $R_0(0) = 1/\sqrt{c_3}$ of the equilibrated most unstable jet with wavenumber $n_c$ as a function of $\beta$.  Dashed lines show the $\beta^{1/3}$ and $\beta^{2/3}$ slopes for reference.  (b) The coefficient $c_3$ and its decomposition into the contributions $c_3^{ec}$ and $c_3^{1,2}$ as a function of $\beta$.  Coefficient $c_3^{1,2}$ is negative for $4.9\lessapprox\beta\lessapprox 79$.  However, for these values of $c_3^{1,2}$ is at least an order of magnitude less than $c_3^{ec}$ and, therefore, negligible.}
\label{fig:R0_c3_gamma0}
\end{figure}

\section{The physical processes underlying the equilibration of the SSD instability of the homogeneous state \label{sec:equil_landau}}

One of the main objectives of this paper is to study the processes that control the halting of the flow-forming instability both for the low branch equilibria, which are governed by the \mbox{G--L} dynamics, and for the upper branch equilibria (cf. Figs.~\ref{fig:nc_ampl1and2} and~\ref{fig:bifurc_b58_jetspectra}).

\subsection{Equilibration processes for the lower branch}

For \mbox{G--L} dynamics, the equilibration of the instability for the most unstable jet perturbation with wavenumber $n_c$ as well as for sideband jets (i.e., jets with scales close to $n_c$) is controlled by coefficient $c_3$ in \eqref{eq:GL}.  We start with a discussion on how $c_3$, and consequently of the equilibration amplitude $R_0(0)$, depends on $\beta$; Fig.~\ref{fig:R0_c3_gamma0}(a) shows the amplitude of the most unstable jet, $R_0(0)$, as a function of~$\beta$.  For $\b\gg1$ the emerging jets have large scales ($n_c\ll1$) and equilibrate with amplitude that increases as $R_0\sim\beta^{1/3}$.  For $\beta\ll1$, the emerging jets have small scales ($n_c\approx 1$) and their amplitude  scales as $R_0\sim\beta^{2/3}$. The scaling of $R_0$ for $\b\gg1$ is found to be robust feature independent of the spectral properties of the forcing (cf. Fig.~\ref{fig:R0_c3_gamma0} and~Fig.~\ref{fig:R0_c3_gamma1}). On the other hand for $\b\gg1$ the amplitude $R_0$ depends crucially on the forcing structure; see \ifdraft Appendix~C\else\hyperref[app:NIF]{Appendix~C}\fi. However, the regime $\b\ll1$ is uninteresting anyway since the anisotropy in the dynamics in~\eqref{eq:NLbarotropic} becomes vanishingly small and no zonal jets emerge.

The dependence of the amplitude $R_0(0)$ on $\beta$ can be understood by considering the contribution of the various wave components to $c_3$, in a similar manner as we did for~$f_r$ in~\eqref{eq:fr}.  Thus, we write:
	\begin{equation}
		c_3 =\int_{0}^{\pi/2} \Fcal_{\rm NL}(\thet)\,\df\thet\ ,\label{eq:NL_anal}
	\end{equation}
where $\Fcal_{\rm NL}$ is the contribution to $c_3$ from the four waves with wavevectors $\kv=(\pm\cos \thet,\pm\sin \thet)$.  Figure~\ref{fig:flux_feedback} shows the contributions $\Fcal_{\rm NL}(\thet)$ for two values of~$\beta$.

For $\beta\ll1$, all wave orientations  contribute positively to~$c_3$.  As a result, the up-gradient contributions to the vorticity flux feedback $\Fcal$ at small $\thet$ are counteracted by $\Fcal_{\rm NL}$, while the down-gradient
contributions to $\Fcal$ at higher $\thet$ are enhanced by $\Fcal_{\rm NL}$.  This leads to a rapid quenching of the instability and thus to a weak finite amplitude jet.

For large~$\beta$, $\Fcal_{\rm NL}$ has roughly the same dipole structure centered about an angle $\thet_0$ as the vorticity flux feedback~$\Fcal$.  Therefore, only waves with angles close to $\thet_0$ contribute appreciably to $c_3$.  Waves with angles $|\thet|<\thet_0$ give positive contributions to $c_3$, while waves with angles $|\thet|>\thet_0$ give negative contributions to $c_3$.  As a result, both the up-gradient and the down-gradient contributions to~$\Fcal$ are almost equally reduced and the instability is only slowly hindered and is allowed to drive jets with a much larger amplitude compared to $\beta\ll1$.  To understand the power law increase of $R(0)$ with $\beta$, note that as $\beta$ increases: \textit{(i)}~the heights of the dipole peaks grow linearly with $\beta$, \textit{(ii)}~the widths of the dipole peaks decrease as $\beta^{-2/3}$, and \textit{(iii)}~the structure of dipole becomes more symmetric about $\thet_0$.  Figure~\ref{fig:FNL_betaF_R2FNL_beta}(a) demonstrates points \textit{(i)}-\textit{(iii)}.  Thus, each of the positive and the negative contribution to $c_3$ scale as $\beta\times\beta^{-2/3}=\beta^{1/3}$ and their difference scales with the derivative, i.e., as $\df\beta^{1/3}/\df\beta \propto \beta^{-2/3}$ leading to the increase of $R(0)$ with $\beta$ as $\beta^{1/3}$.



\begin{figure}
\centerline{\includegraphics[width=19pc]{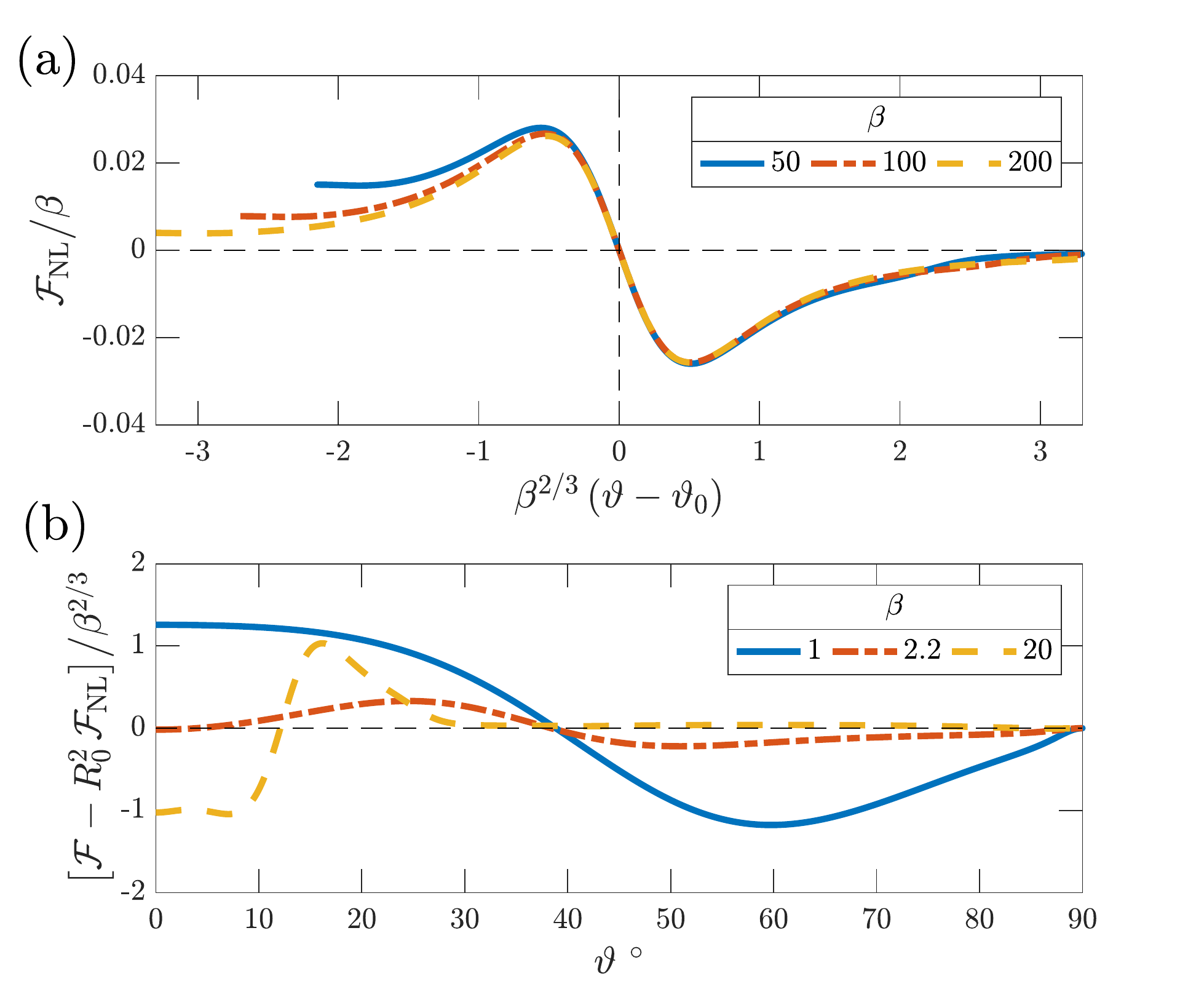}}
\caption{(a) The contribution $\Fcal_{\rm NL}$ to the coefficient $c_3$ from waves at angle $|\thet|$ in the limit of $\beta\gg1$.  $\Fcal_{\rm NL}$
assumes a dipole pattern.  The amplitude of each of the dipole peaks scale with $\beta$ and the widths of the dipole structure scale
with $\beta^{-2/3}$.  For $\beta\gg1$ the structure of $\Fcal_{\rm NL}$ is independent of the type of forcing used.  (b) The contribution from waves at angle $|\thet|$ to the finite amplitude equilibrium jet, as given by $\Fcal-R_0^2\Fcal_{\rm NL}$ for three values of $\beta$.}
\label{fig:FNL_betaF_R2FNL_beta}
\end{figure}

Next we investigate how each of the forced waves contribute in sustaining the equilibrated state of the most unstable jet ($\nu=0$) with amplitude $R_0(0)$ by decomposing the portion of the vorticity flux exceeding dissipation which is the sum of $f_r$ and $-c_3R_0(0)^2$, into contributions from various wave angles:
	\begin{equation}
		f_r-c_3R_0(0)^2=\int_0^{\pi/2} \[\Fcal(\thet) - R_0^2\Fcal_{\rm NL}(\thet)\]\,\df\thet\ .\label{eq38}
	\end{equation}
Figure~\ref{fig:FNL_betaF_R2FNL_beta}(b) shows these contributions for three values of~$\beta$.  For small values of $\beta$ waves with angles $|\thet|<\pi/4$ that drive the instability through their up-gradient contribution also support the equilibrated jet.  However, for $\beta\gg1$ this picture is reversed.  The instability is driven by waves with $|\thet|<\thet_0$ (mainly from waves with $|\thet|\approx 0$) and is hindered by waves with angles $|\thet|>\thet_0$, while the equilibrated jet is supported through the up-gradient fluxes of waves with angles $|\thet|>\thet_0$. The reason is the amplitude $R_0(0)$ is so large that the sign of the integrand in~\eqref{eq38} is reversed.  Further investigation of the eddy--mean flow interactions leading to this peculiar feedback is out of the scope of the current work and will be reported in a future study.


Further insight into the equilibration dynamics is gained by noting that the coefficient $c_3$ can be written as the sum of two separate contributions:
	\begin{equation}
		c_3=c_3^{ec}+c_3^{1,2}\ ,
	\end{equation}
which represent different physical processes (details on the decomposition can be found in \ifdraft Appendix~B\else\hyperref[app:GL]{Appendix~B}\fi).  These contributions correspond
to the two $O(\mu^3)$ possible interactions between the perturbed components of the mean flow $\mu \um_1$ and $\mu^2 \um_2$ with the
covariance corrections $\mu C_1$, $\mu^2 C_{20}$, $\mu^2C_{22}$.

Coefficient
	\begin{equation}
		c_3^{ec} \propto -f\left(0\,|\,\um_1,C_{20}\right)\ ,
	\end{equation}
is proportional to the mean vorticity flux feedback  from the interaction of  $\mu \um_1$ with  the homogeneous covariance correction $\mu^2 C_{20}$ to the equilibrium $C^e$. It measures the compensation in the vorticity flux as perturbations lose energy to the mean flow.

Coefficient
	\begin{equation}
		c_3^{1,2}\propto -f(0 \,|\, \um_1 , C_{22}\,\ee^{2\i n_c (y_a+y_b)/2}+\mbox{c.c.})-f(0\,|\, \um_2 , C_1)\ ,
	\end{equation}
measures the mean vorticity flux feedbacks from the interaction of $\mu \um_1$ and $\mu^2 \um_2$ with the inhomogeneous
covariance corrections $\mu C_{1}$ and $\mu^2 C_{22}\,\ee^{2\i n_c (y_a+y_b)/2}$ to the equilibrium $C^e$.  The exact form of the
coefficients is given in~\eqref{eq:c3_ec} and~\eqref{eq:c3_12} respectively.

Figure~\ref{fig:R0_c3_gamma0}(b) shows the contribution of the two processes in $c_3$ as a function of $\beta$.  We observe that
the main contribution to the coefficient $c_3$ comes from $c_3^{ec}$ for most values of $\beta$.  Only for $\beta\ll1$ is there a
contribution from $c_3^{1,2}$ at the same order.\footnote{Further analysis on the relative contributions of the forced eddies on the
for the two distinct processes can be found in \ifdraft Appendix~B\else\hyperref[app:GLstab]{Appendix~B}\fi.} The
same results also hold for the case of the anisotropic forcing; see Fig.~\ref{fig:R0_c3_gamma1}. 
Therefore, we conclude that for most values of $\beta$, the mean flow is stabilized by the change in the 
homogeneous part of the covariance due to conservation of the total energy that leads to a concomitant
reduction of the up-gradient fluxes. For $\beta\ll1$ there is no change in the eddy--mean flow dynamical processes involved, while for $\beta\gg1$ the equilibrated flow is supported by the up-gradient fluxes of the eddies that were initially hindering its formation.

\subsection{Equilibration processes for the upper-branch jets\label{sec:upper_branch}}

We have seen in the discussion surrounding Fig.~\ref{fig:bifurc_b58_jetspectra}, that the $2n_c$-components of the upper branch equilibria are much stronger than the corresponding $2n_c$-components of the lower branch jets.  Therefore, we expect the interaction between the jet components with wavenumbers $n_c$ and $2n_c$ to play an important role in the equilibration of the upper-branch jets.  This is not at all the case for the lower branch \mbox{G--L} equilibria for which this interaction quantified by $c_3^{1,2}$ is sub-dominant compared to the energy correction term~$c_3^{ec}$.

To investigate the interaction between the jet components with wavenumbers $n_c$ and $2n_c$, we impose a mean flow with power
only at those Fourier components:
	\begin{equation}
		\bar{u}=\uhato\ee^{\i n_c y}-\uhatt\ee^{2\i n_cy}+\mbox{c.c.}\ .\label{eq:flow_adiab}
	\end{equation}
At low supercriticality there is a phase difference of $180\deg$ between the two
components (see~\eqref{eq:2nc} and the fact that $\alpha_2<0$).  Therefore, we impose the same phase difference in~\eqref{eq:flow_adiab}.  We then compute the vorticity fluxes which are induced by the mean flow~\eqref{eq:flow_adiab} by employing the adiabatic approximation, i.e.,
by assuming that the mean flow evolves slow enough that it remains in equilibrium with the eddy covariance and thus $\partial_t C\approx 0$.  Such
an adiabatic approximation is exact for the fixed points of the S3T dynamics but it has also been proven adequate in qualitatively illuminating
the eddy--mean flow dynamics away from the homogeneous or inhomogeneous equilibria
\citep{Farrell-Ioannou-2003-structural,Farrell-Ioannou-2007-structure,Bakas-Ioannou-2013-jas,Bakas-etal-2015}.  With the adiabatic approximation, the Lyapunov equation~\eqref{eq:cov_evo} simplifies to:
	\begin{equation}
		-\Lcal C+\Ncal(\uhato\,\ee^{\i n_c y}-\uhatt\,\ee^{2\i n_c y}+\mbox{c.c},C)+\varepsilon Q=0\label{eq:lyap}\ .
	\end{equation}
We solve~\eqref{eq:lyap} for $C$, we compute the vorticity fluxes and decompose them into their Fourier components:
	\begin{equation}
		\underbrace{\overline{\v'\z'}}_{=\Rcal(C)} = \sum_{m}\hat{f}_{m\,n_c}(\uhato,\uhatt)\ee^{\i m n_c y}+\mbox{c.c.}\ ,
	\end{equation}
with $m$ positive integer.  Then, from the mean flow equation~\eqref{eq:s3tm}, we obtain that the mean flow components satisfy:\begin{subequations}\label{eqs:u1u2tend}
	\begin{align}
		\frac{\df\uhato}{\df t}&=\hat{f}_{n_c}(\uhato,\uhatt)  - \uhato \ ,\\
		\frac{\df\uhatt}{\df t}&=\hat{f}_{2n_c}(\uhato,\uhatt) - \uhatt\ .
	\end{align}\end{subequations}

Figure~\ref{fig:neweq_flux1} shows the mean flow growth rates (e.g.,~$(1/\uhato)\df\uhato/\df t$) as a function of the components $\uhato$ and $\uhatt$ of the imposed mean flow.  We see that for an infinitesimal mean flow (lower left corner of the two panes; noted as region G--L), the growth of $\uhato$ resulting from the linear instability and the growth of $\uhatt$ resulting from the second order self-interaction of the unstable mode (c.f.~\eqref{eq:2nc}) lead to an increase of both $\uhato$ and $\uhatt$.  The flow, thus, equilibrates at the point of intersection of the zero contours for both mean flow tendencies (thick white curves).  This is the lower branch \mbox{G--L} equilibrium that is shown by the open circle and was discussed in the previous section.

There exist, however, two additional points of intersection, both of which are accessible to the flow through paths in the $\uhato$--$\uhatt$ parameter space.  If we start with a strong $\uhato\gtrapprox 0.14$ component from point~A in the figure, the large positive growth rate $(1/\uhatt)\df\uhatt/\df t$ will lead to a rapid increase of $\uhatt$, while the slightly negative tendency $(1/\uhato)\df\uhato/\df t$ will gradually weaken $\uhato$ so that $\uhatt$ and $\uhato$ will move towards the right point of intersection.  We perform an integration of the S3T dynamical system \eqref{eq:s3ts} with initial conditions starting from point~A.  The path of the dynamical system in the $\uhato$--$\uhatt$ parameter space that is shown by the dotted line, confirms the qualitative picture obtained via the mean flow growth rates with the rapid increase of $\uhatt$ and the eventual equilibration at the right point of intersection shown by the filled triangle.  Similarly, if we start with a strong $\uhatt\gtrapprox 0.2$ component from point~B, the strong growth $(1/\uhato)\df\uhato/\df t$ and the weak negative tendency $(1/\uhatt)\df\uhatt/\df t$ lead again to the equilibration of the flow through the path shown in Fig.~\ref{fig:neweq_flux1}.  The growth rates close to the other point of intersection shown by the open circle reveals that this corresponds to an unstable equilibrium and this is also confirmed through integrations of the S3T system~\eqref{eq:s3ts}.  These two points therefore correspond to the stable and unstable equilibria of the upper branch that are shown in Fig.~\ref{fig:bifurc_b58_jetspectra}.

The qualitative agreement between the approximate dynamics of~\eqref{eqs:u1u2tend} and the nonlinear S3T dynamics reveal that it could be a useful tool for exploring the phase space of the S3T system.  For example, the bifurcation structure of Fig.~\ref{fig:bifurc_b58_jetspectra} could be obtained by plotting the adiabatic growth rates.  Figure~\ref{fig:neweq_flux2} shows the curves of zero tendencies for various values of the supercriticality.  For low subcritical values
$\e/\ec<0.89$ (panel (a)), there is no point of intersection, therefore only the homogeneous equilibrium exists.  For $0.89\leq \e/\ec<1$ (panel b), there are
two points of intersection revealing the existence of the stable and the unstable upper branch equilibria, while for $1\leq\e/\ec$ (panel (c)) there is the
additional lower branch point.  Finally, for highly supercritical values (panel (d)) there is only one point of intersection revealing the existence of
the stable upper branch equilibrium.

\begin{figure*}
\centerline{\includegraphics[width=33pc]{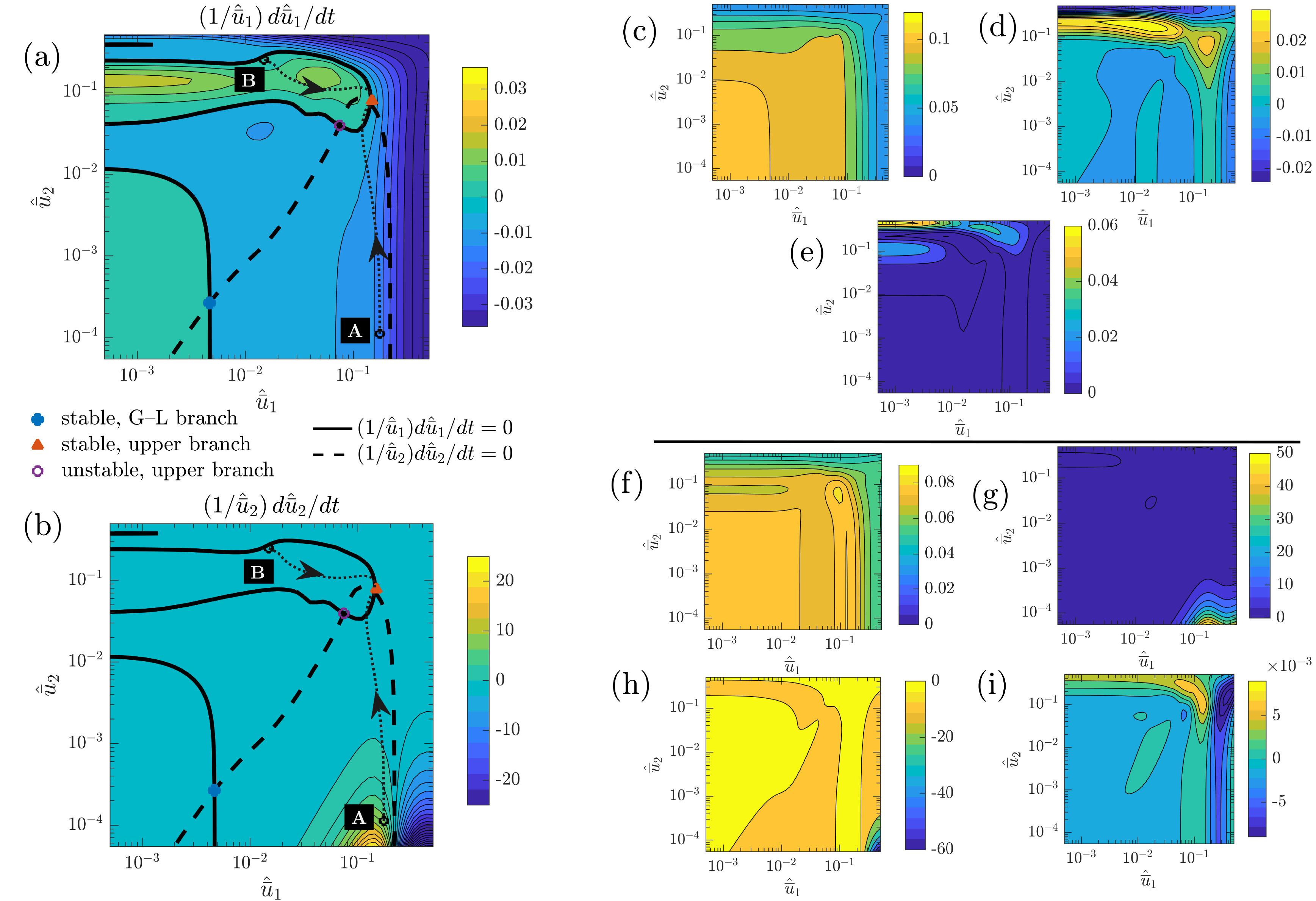}}
\caption{The mean flow growth rates $(1/ \uhato)\df\uhato/\df t$ (panel (a)) and $(1/\uhatt)\df\uhatt/\df t$ (panel (b)) obtained under an adiabatic approximation ($\partial_tC=0$) for a mean flow $\bar{u}=\uhato\ee^{\i n_cy}-\uhatt\ee^{2\i n_c y}$ as a function of $\uhato$ and $\uhatt$.  The thick curves are the
zero tendency contours (solid curve for the $\df\uhato/\df t$ and dashed for $\df\uhatt/\df t$).  Infinitesimal jet perturbations start in the region in the
$\uhato$--$\uhatt$ phase space denoted as \mbox{G--L} and end up in the lower branch equilibrium shown by the filled circle.  The arrows denote
paths in the $\uhato$--$\uhatt$ phase space that connect a finite amplitude jet perturbation starting from points~A and~B and ending up to the upper branch equilibrium, denoted by the filled triangle.  (The paths were obtained by time-stepping the S3T system~\eqref{eq:s3ts}.) Panels~(c)--(e) show the breakdown of the flux feedback $\hat{f}_{n_c\kfd}/\uhato$ into the components (c)~$f_{1,0}/\uhato$, (d)~$f_{1,2}/\uhato$ and
(e)~$f_{2,3}/\uhato$.  Similarly, panels~(f)--(i) show the breakdown of $\hat{f}_{2n_c\kfd}/\uhatt$ into the components (f)~$f_{2,0}/\uhatt$,
(g)~$f_{1,1}/\uhatt$, (h)~$f_{1,3}/\uhatt$ and (i)~$f_{2,4}/\uhatt$.  Parameters used: $\beta_{58}$ and supercriticality $\mu=0.1$.}
\label{fig:neweq_flux1}
\end{figure*}

\begin{figure*}
\centerline{\includegraphics[width=34pc]{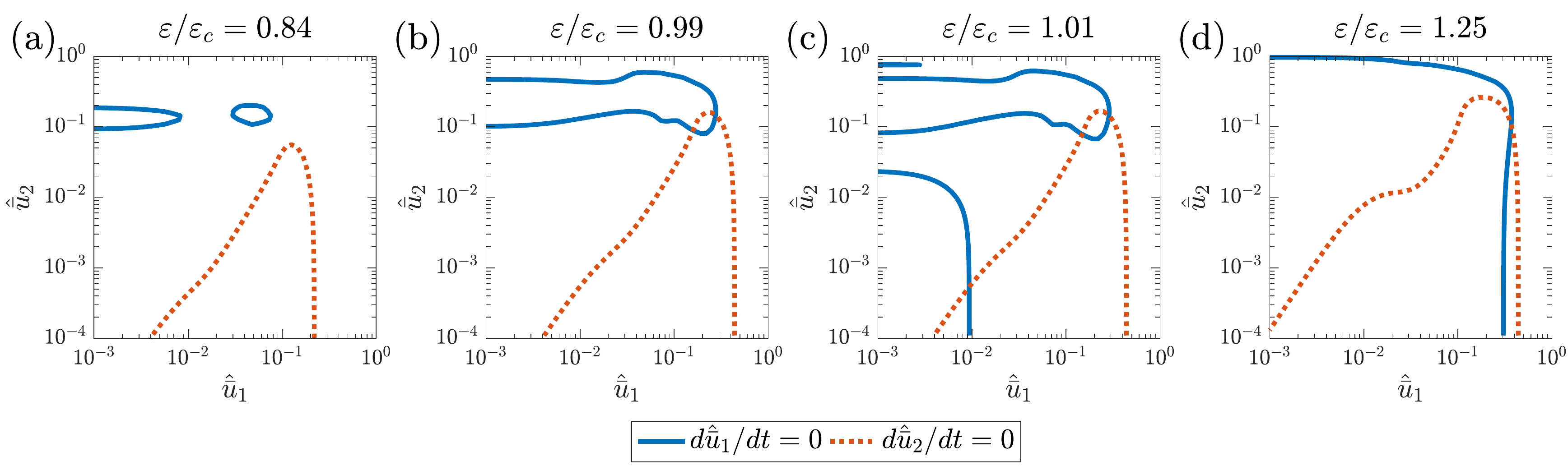}}
\caption{The locus of zero mean flow tendencies in $\uhato$--$\uhatt$ space for various supercriticalities for the case
with $\beta_{58}$.  Jet equilibria exist at the intersection of the two loci, when $d\uhato/dt=d\uhatt/dt=0$.  For (a)
$\e/\ec=0.84$ there is no intersection as only the homogeneous equilibrium is stable for $\e/\ec<0.89$ (see
Fig.~\ref{fig:bifurc_b58_jetspectra}(a)).  For (b) $\e/\ec=0.99$ there are two points of intersection that correspond to the
stable and unstable upper branch equilibria that exist for $0.89<\e/\ec<1$.  For (c) $\e/\ec=1.01$ there are three equilibria:
two upper-branch equilibria (a stable and an unstable) and the lower-branch \mbox{G--L} equilibrium.  For (d) $\e/\ec=1.25$ only
one upper branch equilibrium exists as the \mbox{G--L} branch and the unstable upper branch terminate at $\e/\ec=1.068$.}
\label{fig:neweq_flux2}
\end{figure*}

To shed light into the dynamics underlying these new equilibration paths that lead to the upper-branch
equilibria, we decompose the covariance as a Fourier sum over the inhomogeneous components
\begin{equation}
	C=\sum_{m=0}^4\Chat_{m}(\bx_a-\bx_b)\,\ee^{\i m n_c (y_a+y_b)/2}\ .
\end{equation}
The sum is over five components.  The reason is that first of all the flux feedback on $\uhato$ and $\uhatt$ is generated by the
$n_c$ and $2n_c$ components of the covariance.  Inspection of the nonlinear term in~\eqref{eq:lyap},
reveals that only the homogeneous component $\Chat_0$ as well as the covariance components at $n_c$, $2n_c$, $3n_c$
and $4n_c$ can interact with the mean flow~\eqref{eq:flow_adiab} to yield these two covariance components.  We then decompose
the vorticity fluxes as:
\begin{subequations}
\begin{align}
	\hat{f}_{n_c}(\uhato,\uhatt)&=f_{1,0}+f_{1,2}+f_{2,3}\ ,\label{eq:flux_feed1}\\
	\hat{f}_{2n_c}(\uhato,\uhatt)&=f_{2,0}+f_{1,1}+f_{1,3}+f_{2,4}\ .\label{eq:flux_feed2}
\end{align}\label{eqs:flux_feeds}\end{subequations}
Each of the terms on the right-hand-sides of~\eqref{eqs:flux_feeds} represents the different interactions among
the mean flow components $\uhato$, $\uhatt$ with the covariance components $\Chat_0$, $\Chat_1$, $\Chat_2$, $\Chat_3$ and $\Chat_4$.
The first term in~\eqref{eq:flux_feed1} is proportional to the vorticity flux feedback from the interaction of $\uhato$ with the
homogeneous covariance component $\Chat_0$:
\begin{equation}
f_{1,0}\propto f\left(0\,|\,\uhato\ee^{\i n_c y},\Chat_0\right).
\end{equation}
For low supercriticality,
\begin{equation}
f_{1,0}-\uhato \approx A\left(1-c_3^{ec}|A|^2\right).
\end{equation}
This means that $f_{1,0}$ contains both the destabilizing feedback which drives the linear instability, and the stabilizing feedback at finite amplitude that results from the energy correction.  The terms $f_{1,2}$ and $f_{2,3}$ in~\eqref{eq:flux_feed1} describe the feedback of the nonlinear interaction between $\uhato$ and $\uhatt$ on $\uhato$:
	\begin{align}
		\ifdraft
		f_{1,2}&\propto f ( 0\,|\,\uhato^*\ee^{-\i n_cy},\Chat_2\ee^{2\i n_c(y_a+y_b)/2} )  +f ( 0\,|\,\uhatt\ee^{2\i n_cy},\Chat_1^*e^{-\i n_c(y_a+y_b)/2} )\ ,\\
		\else
		f_{1,2}&\propto f ( 0\,|\,\uhato^*\ee^{-\i n_cy},\Chat_2\ee^{2\i n_c(y_a+y_b)/2} ) \nonumber\\
				& \qquad +f ( 0\,|\,\uhatt\ee^{2\i n_cy},\Chat_1^*e^{-\i n_c(y_a+y_b)/2} )\ ,\\
		\fi
		f_{2,3}&\propto f\left(0\,|\,\uhatt^*\ee^{-2\i n_cy},\Chat_3\ee^{3\i n_c(y_a+y_b)/2}\right)\ .
	\end{align}
For low supercriticality
	\begin{equation}
		f_{1,2}\approx-c_3^{1,2}A|A|^2\ ,
	\end{equation}
while $f_{2,3}$ is of higher order in $\mu$.  Similarly, the second term on the right-hand-side of~\eqref{eq:flux_feed2} is
proportional to the vorticity flux feedback from the interaction of $\uhatt$ with
the homogeneous covariance $\Chat_0$:
	\begin{equation}
		f_{2,0}\propto f ( 0\,|\,\uhatt\,\ee^{2\i n_c y}, \Chat_0 )\ .
	\end{equation}
For low supercriticality, the flux feedback above is positive but does not overcome friction, i.e.,
$0<f_{2,0}<\uhatt$.  Therefore, the homogeneous equilibrium is linearly stable with respect to jet perturbations with wavenumber
$2n_c$ (as expected).  The terms $f_{1,1}$, $f_{1,3}$ and $f_{2,4}$ in~\eqref{eq:flux_feed2} describe the feedback of the
nonlinear interaction between $\uhato$ and $\uhatt$ on $\uhatt$:
	\begin{align}
		f_{1,1} &\propto f\left(0\,|\,\uhato\ee^{\i n_cy}, \Chat_1\ee^{\i n_c(y_a+y_b)/2}\right)\ ,\\
		f_{1,3} &\propto f\left(0\,|\,\uhato^*\ee^{-\i n_cy}, \Chat_3\ee^{3\i n_c(y_a+y_b)/2}\right)\, \\
		f_{2,4} &\propto f\left(0\,|\,\uhato^*\ee^{-2\i n_cy}, \Chat_4\ee^{4\i n_c(y_a+y_b)/2}\right)\ .
	\end{align}

For low supercriticality, $f_{1,1}$ drives the $\uhatt$ component of the flow
with an amplitude proportional to $\uhato^2$ and, therefore, $\uhatt$ equilibrates at amplitude~\eqref{eq:2nc}, while $f_{1,3}$ and
$f_{2,4}$ are of higher order.
Panels (c)-(i) of Fig.~\ref{fig:neweq_flux1} show the contribution of the various terms to the flux feedbacks $\hat{f}_{n_c}$ and
$\hat{f}_{2n_c}$ respectively.  In the \mbox{G--L} region the fluxes are determined by $f_{0,1}$, $f_{0,2}$ and $f_{1,1}$.  However,
the ``tongue'' of positive tendency $(1/\uhato)\df\uhato/\df t$ in Fig.~\ref{fig:neweq_flux1}(a) for large values of $\uhatt$, as well as the
region of very large positive tendency $(1/ \uhato)\df\uhato/\df t$ in Fig.~\ref{fig:neweq_flux1}(b) are determined by the other terms.  As a result,
the equilibration of the flow in the upper layer branch is due to the nonlinear interaction of the two mean flow components $\uhato$
and $\uhatt$ rather than the energy correction that underlies the equilibration of the flow in the lower branch.

\section{Eckhaus instability of the side band jets\label{sec:Eckhaus}}

In this section we study the stability of the sideband jet equilibria.  As noted by~\cite{Parker-Krommes-2014-generation}, these
harmonic jet equilibria are susceptible to Eckhaus instability, a well known result for harmonic equilibria of the \mbox{G--L}
equation \citep{Hoyle-2006}.  Here, we present the main results of the Eckhaus instability and compare them with fully nonlinear
S3T dynamics.

\subsection{An intuitive view of the Eckhaus instability}

To obtain intuition for the eddy--mean flow dynamics underlying the Eckhaus
instability, note first that the \mbox{G--L} dynamics are given by the balance between the vorticity flux
feedback $f_r(\nu)=f_r(0)-c_2\nu^2$, which  provides a diffusive correction to the
original up-gradient fluxes $f_r(0)>0$ at $n_c$, and the stabilizing nonlinear term $c_3|A|^2$.  Let us assume an
equilibrium jet with $\nu>0$, i.e.  with a
scale smaller than that of the most unstable jet at $n_c$, and also assume a sinusoidal phase perturbation:
	\begin{equation}
		A(Y)=R_0\,\ee^{\i[ \nu Y+\eta \sin(qY)]}\quad\textrm{with}\quad \eta\ll1\ .\label{eq43}
	\end{equation}
Figure~\ref{fig:landau_Eckhaus} shows how the perturbed jet~\eqref{eq43} is compressed for half the wavelength of the
phase perturbation~$\pi/q$ (unshaded region) and dilated for the other half (shaded region).  In the compressed region the
jet appears with an enhanced wavenumber $\nu+\delta \nu$ while in the dilated region the jet appears with a reduced wavenumber
$\nu-\delta\nu$.  As a result, the vorticity flux feedback $f_r(\nu)$ is larger in the dilated (shaded) region implying a tendency
to enhance the jet; the opposite occurs in the compressed region (non-shaded).  Figure~\ref{fig:landau_Eckhaus} shows a qualitative
sketch of the mean vorticity fluxes, $\overline{\v'\zeta'}$, that demonstrates this process.  If the nonlinear term does not counteract
this mismatch, the dilated part of the jet will grow and take over the whole domain thus producing a jet with lower $\nu$. (Similarly, for an equilibrium jet with $\nu<0$ there is a tendency for the compressed part of the jet to take over the whole domain producing a jet with larger $\nu$.)

To summarize, due to the diffusive nature of the vorticity flux feedback there is a tendency to go towards $\nu=0$ jets if
not counteracted by the nonlinear eddy--mean flow feedback.

\begin{figure}[ht]
\centerline{\includegraphics[width=9pc]{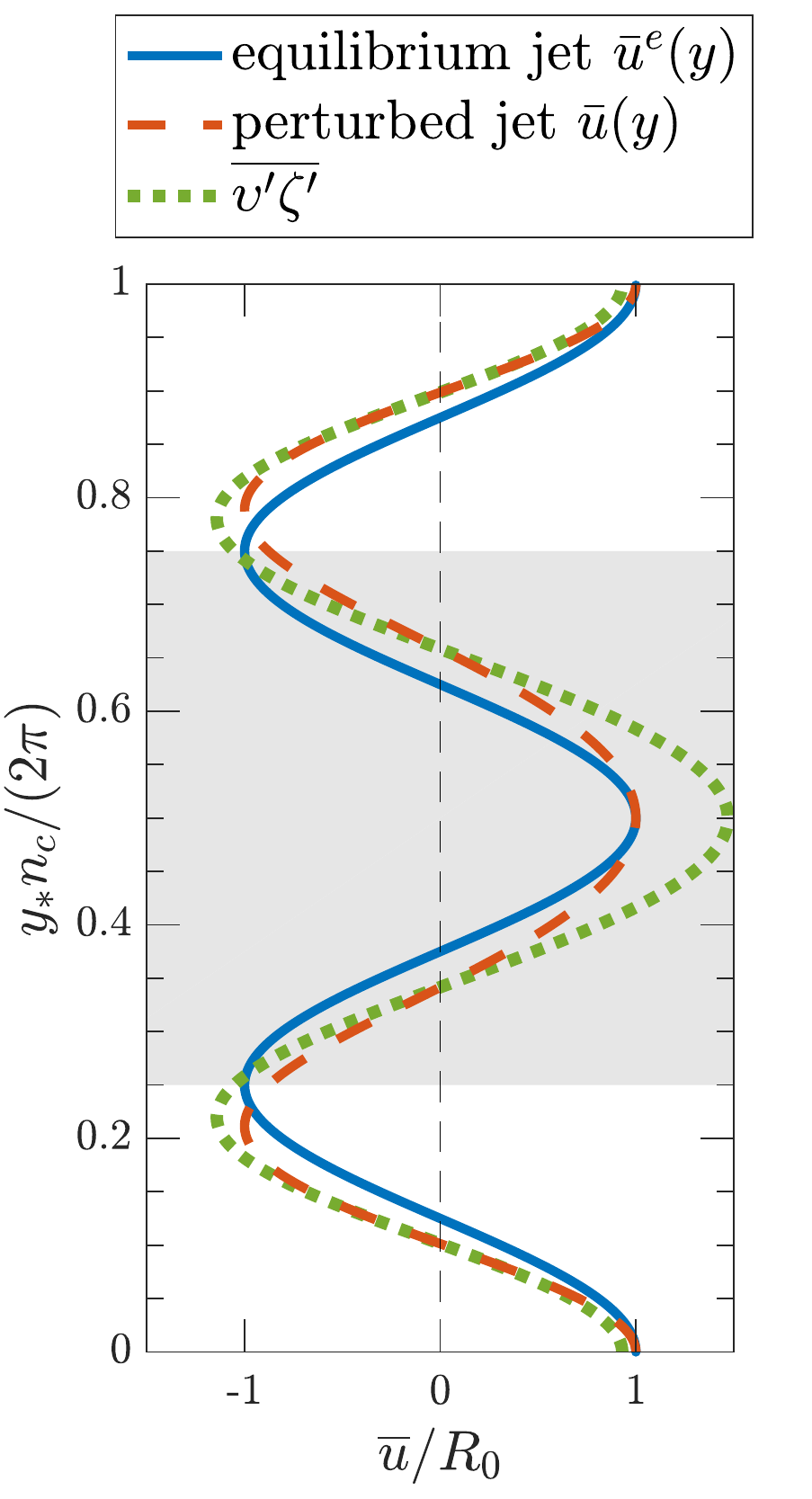}}
\vspace{-1em}\caption{Solid curve shows a sinusoidal equilibrium jet $\um^e=R_0\cos\left[(n_c+\mu\nu) y\right]$ with smaller scale ($\nu=n_c$) compared to the scale of the
most unstable jet  (we take $\mu=1$ so that the wavenumber differences with the most unstable jet are exaggerated for illustration purposes).
Dashed curve shows the resulting jet when the phase of the equilibrium jet $\um^e$ is perturbed:
$\um=R_0\cos\left[(n_c+\mu\nu) y+\eta\sin(qy)\right]$, with $q=\nu$ and $\eta=1/2$ (for illustration purposes).  This perturbation dilates the jet
in the shaded region and compresses the jet in the unshaded region. Dash-dotted curve is a qualitative depiction of the expected vorticity flux feedback
$\overline{\v'\zeta'}$ for the perturbed jet based on the dependence of $f_r$ on the wavenumber $\nu$.}
\label{fig:landau_Eckhaus}
\end{figure}

\subsection{A formal view of the Eckhaus instability}

To address quantitatively the stability of the harmonic jet equilibria~\eqref{eq:eq_amp}, let us reformulate
the \mbox{G--L} equation by rewriting the jet amplitude $A$ in polar form as:
	\begin{equation}
		A(Y,T)=R(Y,T)\,\ee^{\i\Theta(Y,T)}\ ,\label{def_A_thet}
	\end{equation}
where $R$ is the amplitude and $\Theta$ is the phase of the jet.  The equilibrium jets have a constant amplitude
$R_0(\nu)$ given by~\eqref{eq:eq_amp} and a linearly varying phase $\Theta=\nu Y$.  From~\eqref{eq:nucrit}, such
equilibria exist only for $|\nu|<\nu_e=\sqrt{1/c_2}$.  Consider now small perturbations about this equilibrium jet:
	\begin{equation}
		R = R_0(\nu)+\hat{\rho}\,\ee^{iqY+\lambda T}\quad\text{and}\quad \Theta=\nu Y+\hat{\phi}\,\ee^{iqY+\lambda T}\ .
	\end{equation}
As shown in \ifdraft Appendix~D\else\hyperref[app:GLstab]{Appendix~D}\fi, we have exponential growth of these perturbations
if
	\begin{equation}
		q^2+2(\nu_e^2-3\nu^2)<0.\label{eq:Eck_crit}
	\end{equation}
For an infinite domain the gravest mode has $q=0$ and therefore the jets with amplitude~\eqref{def_A_thet} are Eckhaus unstable
when $|\nu| > \nu_e/\sqrt{3}$.  Maximum instability occurs for
	\begin{equation}
		|q|_{\mbox{max}}=\nu_e\frac{\sqrt{3(\nu/\nu_e)^4+2(\nu/\nu_e)^2-1}}{2(\nu/\nu_e)}\ ,
	\end{equation}
and therefore, the Eckhaus instability will form a jet of wavenumber $n_c +\mu(\nu\pm |q|_{\mbox{max}})$.
Figure~\ref{fig:eckhaus_comp}(a) shows the wavenumber $|q|_{\mbox{max}}$ as a function of the equilibrium jet wavenumber $\nu$.  Note that the equilibria with wavenumbers $\nu\approx \nu_e/\sqrt{3}$ are unstable to jets with neighboring wavenumbers as $|q|_{\mbox{max}}\ll1$, while equilibria with wavenumbers $\nu\approx \nu_e$ are unstable to the jet with wavenumber $n_c$ as $|q|_{\mbox{max}}\approx 1$.

The growth rate for the most unstable structure with $|q|_{\mbox{max}}$ is
	\begin{equation}
		\lambda_{\mbox{max}}=\frac{(3\nu^2-\nu_e^2)^2}{4c_1\nu_e^2\nu^2}\ . \label{eq:maxgreck}
	\end{equation}
and is shown in Fig.~\ref{fig:eckhaus_comp}(b).

\subsection{Comparison with S3T dynamics}

Compare first the stability analysis for the harmonic jets derived in the weakly nonlinear limit of \mbox{G--L}
dynamics to nonlinear dynamics in the S3T system.  Note that the growth rate of the Eckhaus instability is much less than the corresponding growth rate of the flow-forming instability of the
homogeneous state of a jet for almost all wavenumbers $\nu$. Figure~\ref{fig:eckhaus_comp}(b) compares the growth rate $\lambda_{max}$ for the perturbation with $|q|_{max}$ that will eventually form a jet with wavenumber $n_c +\mu(\nu\pm |q|_{\mbox{max}})$ to the growth rate of the flow-forming instability of the homogeneous equilibrium that will form a jet with the same wavenumber (shown with dashed line). As a result, the weak Eckhaus instability manifests only in carefully contrived S3T simulations; any simulation of the S3T system~\eqref{eq:s3ts} starting from a random initial perturbation at low supercriticality will evolve into the most unstable jet with wavenumber $n_c$.

Second, in contrast with the infinite domain, for the doubly periodic box the first side band jets appear when $\nu\geq \nu^\pm$, while the gravest wavenumber $q$ is $q_{\mathrm{min}}\defn 1/(\mu\kfd\Ld)$.  Therefore, the instability criterion~\eqref{eq:Eck_crit} is satisfied for
	\begin{equation}
		\mu\leq \frac{\sqrt{\left(3-1/2\right)c_2}}{\kfd\Ld}.\label{eq:stab_bound}
	\end{equation}
We compare here the stability boundary~\eqref{eq:stab_bound} with the stability analysis based on the nonlinear S3T dynamics. The stability of the inhomogeneous jet--turbulence S3T equilibria shown in Fig.~\ref{fig:sideband_eq} is studying using the numerical methods developed by \citet{Constantinou-2015-phd, Constantinou-etal-2016}; for the stability boundary~\eqref{eq:stab_bound} we use the effective values $c_2^{\mathrm{ex}\pm}$ for the side-band jet equilibria with $\nu^\pm$.  Unstable (stable) equilibria are shown in Fig.~\ref{fig:sideband_eq} with open (filled) symbols, while the stability boundaries for $\nu^\pm$ are shown with the vertical dotted lines.  For $\beta_6$, the parabolic profile of the eigenvalue relation, on which the Eckhaus instability calculations are based, remains accurate for larger supercriticalities and, therefore, the stability boundary~\eqref{eq:stab_bound} consists a good approximation.  For larger and smaller values of $\beta$, the parabolic profile is not so accurate and, therefore, the criterion developed fails.  For example, for both $\beta_1$ and $\beta_{192}$ all the $\nu^+$ jet equilibria are unstable.


\begin{figure}[ht]
\centerline{\includegraphics[width=19pc]{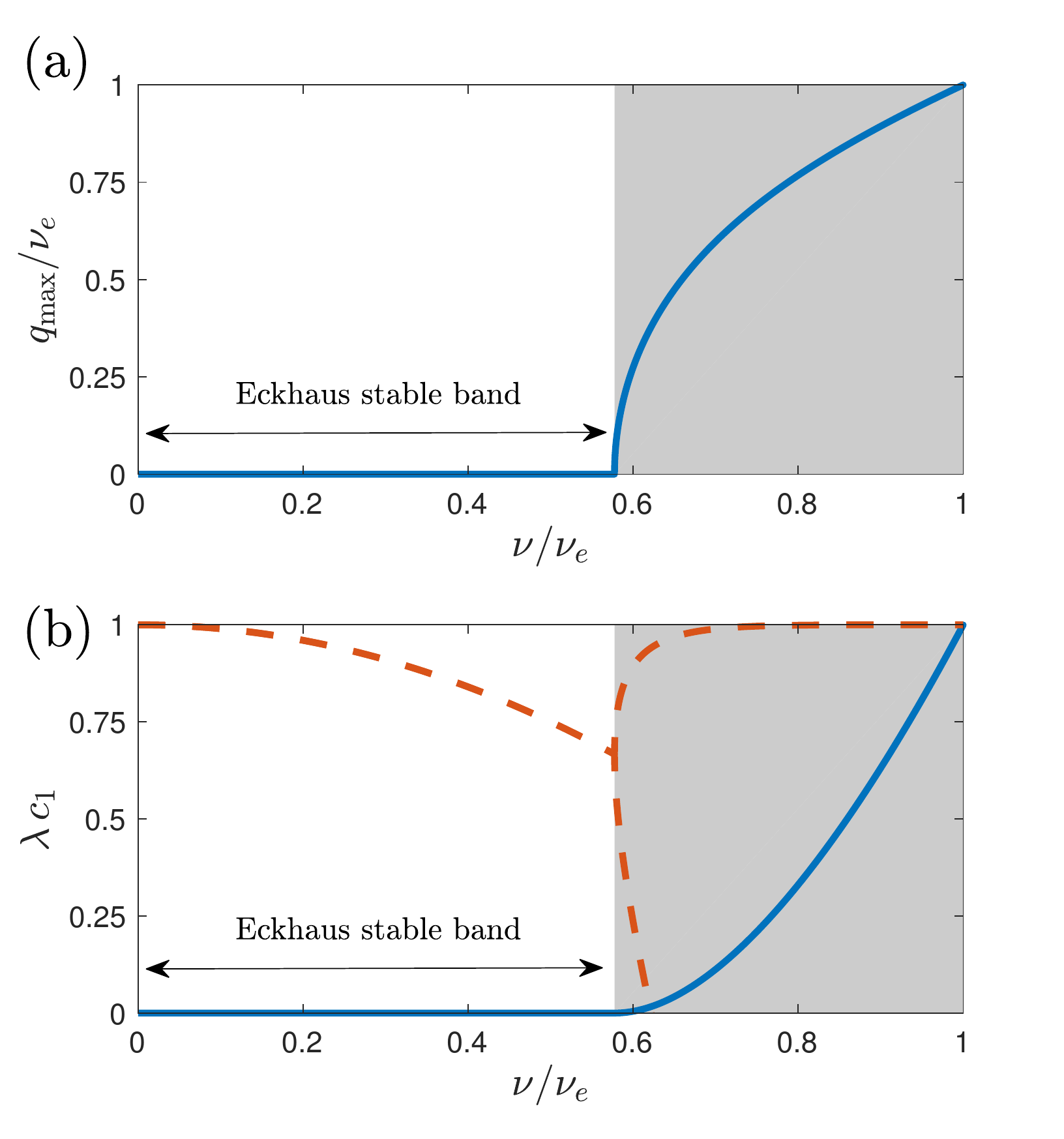}}
\vspace{-1em}\caption{(a) The most unstable wavenumber for the Eckhaus instability, $|q|_{\mbox{max}}/\nu_e$, as a function of
the jet equilibrium wavenumber $\nu/\nu_e$.  Instability occurs in the shaded region for $\nu/\nu_e>1/\sqrt{3}$. (b) The growth rate for
the most Eckhaus unstable jet with $q=q_{\mbox{max}}$~\eqref{eq:maxgreck} as a function of the jet equilibrium wavenumber $\nu$ (solid line).
Also shown with dashed line is the corresponding growth rate for the flow-forming instability of the jet with wavenumber $\nu\pm q_{\mbox{max}}$ that will eventually be formed by the Eckhaus instability and is given by $\left[1-(\nu\pm q_{\mbox{max}})^2/\nu_e^2\right]/c_1$, according to the \mbox{G--L} equation~\eqref{eq:GL}.}
\label{fig:eckhaus_comp}
\end{figure}

\begin{figure*}
\centerline{\includegraphics[width=0.9\textwidth]{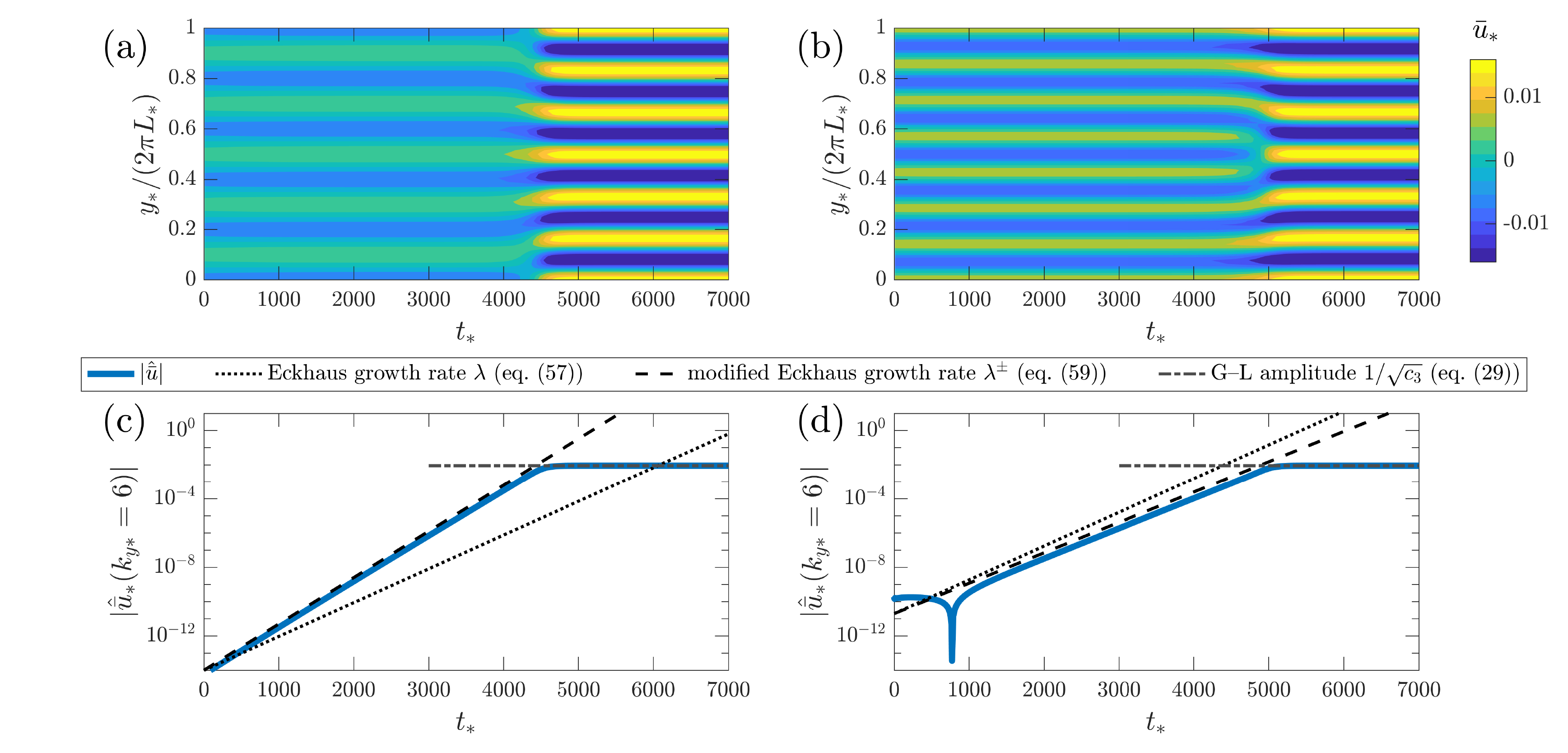}}
\vspace{-1em}\caption{The equilibration of the Eckhaus instability under S3T dynamics.  Panel (a) shows the evolution of the mean flow, $\um_*(y,t)$, for the slightly perturbed $n_*=5$ ($\nu^-$) equilibrium.  Panel (c) shows the evolution of the $n_*=6$ Fourier component of the flow (solid).  Also shown in panel (b) are the growth rate predicted by~\eqref{eq:growth_side} (dashed) and the amplitude of the $n_*=6$ jet as predicted by~\eqref{eq:eq_amp0} (dash-dot).  Panels (b) and~(d) show the same but for the slightly perturbed $n_*=7$ ($\nu^+$) equilibrium.  The planetary vorticity gradient is $\beta_6$ and the supercriticality is $\mu=0.3$.} \label{fig:Eckhaus2}
\end{figure*}

Last, we compare the development of the Eckhaus instability as predicted by the \mbox{G--L} dynamics~\eqref{eq:GL}
and as predicted by the S3T dynamics.  Figure~\ref{fig:Eckhaus2} shows the evolution of the slightly perturbed $n_*=5$ ($\nu^-$) and $n_*=7$ ($\nu^+$) equilibria for $\beta_6$ and supercriticality $\mu=0.3$ obtained from integrations of the S3T system~\eqref{eq:s3ts}.  In both cases, the equilibria are unstable to $q=q_{\mathrm{min}}$ perturbations.  As the instability develops the $\hat{\bar{u}}(k_{y*}=6)$ component of the flow grows exponentially (panels (c) and (d)) and the flow moves into the stable $n_*=6$ ($n_c$) equilibrium jet by branching or merging (panels~(a) and~(b)).  We compute the growth rate of the Eckhaus instability from~\eqref{eq:Eck_growth} by substituting $\nu=q=1/(\mu\kfd\Ld)$ and using the effective values $c_2^{\mathrm{ex}\pm}$:
\begin{equation}
	\lambda^\pm=\mu^2\frac{-1+\sqrt{\left[(\mu^{\mathrm{ex}\pm}/\mu)^2-1\right]^2+4(\mu^{\mathrm{ex}\pm}/ \mu)^4}}{c_1}.\label{eq:growth_side}
\end{equation}
Panels~(c) and~(d) demonstrate that the growth rate obtained by~\eqref{eq:growth_side} is in excellent agreement with the growth rate of the Eckhaus instability in the nonlinear simulations.  Furthermore, the equilibrium jet amplitude is accurately predicted by~\eqref{eq:eq_amp0}.

Figure~\ref{fig:Eckhaus3} shows the comparison of the growth rates for the other unstable sideband jet equilibria illustrated in Fig.~\ref{fig:sideband_eq}. We see once more that for $\beta_6$, for which the parabolic approximation of the eigenvalue relation used to obtain the \mbox{G--L} dynamics is accurate, the growth rates agree for almost all the unstable range.  For $\beta_1$ and $\beta_{192}$, for which the parabolic profile is not accurate, there is in general disagreement.

\begin{figure*}
\centerline{\includegraphics[width=0.9\textwidth]{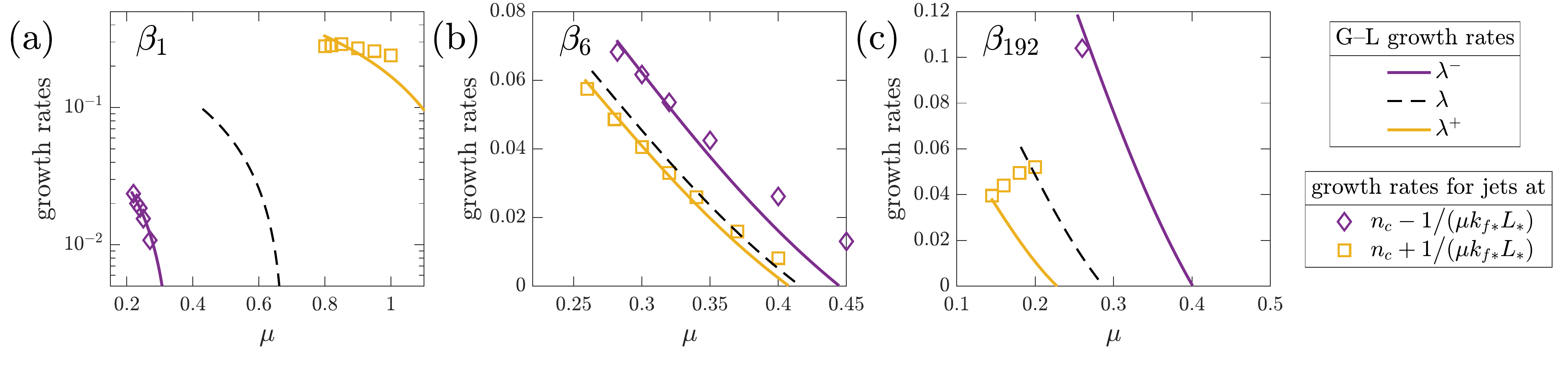}}
\vspace{-1em}\caption{Growth rate for the Eckhaus instability of the finite amplitude jets. Shown is the growth rate as a function
of supercriticality $\mu$ for three values of $\beta$ obtained from the stability analysis for the equilibrium jets
with wavenumbers $n_c-1/(\mu k_{f*}L_*)$ (diamonds) and $n_c+1/(\mu k_{f*}L_*)$ (squares) using the fully nonlinear system~\eqref{eq:s3ts}.
Dashed curves show the growth rate as predicted from the \mbox{G--L} dynamics;
Solid curves show the growth rate~\eqref{eq:growth_side} as predicted from the \mbox{G--L} dynamics using the modified values
for $c_2$, while dashed curves show the unmodified growth rate~~\eqref{eq:Eck_growth}.}\label{fig:Eckhaus3}
\end{figure*}

\section{Conclusion \label{sec:concl}}

We examined the dynamics that underlies the
formation and support of zonal jets at finite amplitude in forced--dissipative
barotropic beta-plane turbulence using the statistical state dynamics of the turbulent flow closed at second-order.  Within this framework, jet formation is shown to arise as a flow-forming instability (or `zonostrophic instability') of the homogeneous statistical equilibrium turbulent state when the non-dimensional parameter~$\e = \ed/(\kfd^{-2}\rd^3)$ crosses a certain critical threshold~$\e_c$.  In this work, we studied the dynamics that govern the equilibration of the flow-forming instability in the limit of small supercriticality $\mu=\sqrt{\varepsilon/\varepsilon_c-1}$.

When supercriticality $\mu\ll 1$, the growth rate of the unstable modes as a function of the mean flow wavenumber is to a good approximation a parabola.  This allows a two-time, two-scale approximation of the nonlinear dynamics resulting in the weakly nonlinear Ginzburg--Landau dynamics for the evolution of zonal jets. The equilibration of the flow-forming instability, was extensively investigated using the \mbox{G--L} dynamics.
Also, the predictions of the weakly nonlinear \mbox{G--L} dynamics regarding \emph{(i)}~the amplitude of the equilibrated jets and \emph{(ii)}~their stability were compared to the fully nonlinear S3T dynamics for a wide range of values for the non-dimensional parameter $\b=\betad/(\kfd\rd)$.

According to \mbox{G--L} dynamics, the harmonic unstable modes of the homogeneous equilibrium state equilibrate at finite amplitude. The predicted amplitude of the jet that results from the equilibration of the most unstable mode with wavenumber $n_c$, was compared to the amplitude of the jet equilibria of the nonlinear S3T dynamics.  For $\beta\lessapprox20$, the jet amplitude was found to be accurately predicted by the \mbox{G--L} dynamics for up to $\mu\approx 0.2$.  For $\beta\gtrapprox20$, a new branch of jets with much larger amplitudes
was discovered that was distinctly different from the \mbox{G--L} branch of jet equilibria.  The bifurcation diagram (e.g.,~Fig.~\ref{fig:bifurc_b58_jetspectra}) exhibits a classic  cusp bifurcation with hysteretic loops. The new  branch of jet equilibria exists even at  \emph{subcritical values}
of  the flow-forming instability of the homogeneous state (i.e., for $\e<\e_c$).  This has two consequences: first, continuation methods  for  finding equilibria
converge only for  small supercriticalities, as  the jet equilibria transition discontinuously to the upper  branch (see,~e.g.,~Fig.~\ref{fig:bifurc_b58_jetspectra}(a)). This explains the failure to converge to equilibria reported by
 \citet{Parker-Krommes-2014-generation}.
  Second, the cusp bifurcation allows
the emergence of jets at subcritical parameter values through a nonlinear flow-forming instability.

We compared the amplitudes of the jets that emerge from the side-band jet-instabilities of the most unstable mode of the flow-forming instability (i.e., the jets that emerge at scales $\ncd \pm 1/\Ld$). The amplitude predicted by the \mbox{G--L} equation is partially based on the parabolic approximation to the dispersion relation and, more specifically, on the curvature of the function of the growth rate at criticality.  This approximation was found to be valid away
from criticality only for non-dimensional $\beta\approx 5$ and as a result the predicted amplitude fails outside
this range.  We propose a way to remedy this discrepancy (at least to some extend) by using the exact values for the curvature of the growth rate function for larger supercriticalities instead of the curvature given by the parabolic approximation (see, e.g., Fig.~\ref{fig:Eckhaus2}).  With this modification, the side-band jet amplitudes can be predicted by the \mbox{G--L} dynamics close to their onset for $\beta\lessapprox1$ and for a wide range of supercriticalities for $\beta\approx 5$.  For $\beta\gtrapprox20$, apart from the \mbox{G--L} branch the additional branch of higher amplitude side-band jets was also found.

The physical and dynamical processes underlying the equilibration of the flow-forming instability were then examined using three
methods.  The first was the decomposition of the nonlinear term in the \mbox{G--L} equation governing the equilibration of the
instability in two terms.  One involves the change in the homogeneous part of the eddy covariance that is required by total energy
conservation.  The other involves the vorticity flux feedback resulting from the interaction of
the most unstable jet with wavenumber~$n_c$ and the jet with the double harmonic~$2n_c$ that is inevitably generated by the nonlinear interactions.  The second was the method of \citet{Bakas-etal-2015} for separating
the contributions of the various eddies in the induced vorticity fluxes: both for the linear
term in the \mbox{G--L} equation that drives the instability, and also for the nonlinear term that stabilizes the flow.  In this way, the
eddies yielding up-gradient fluxes and the eddies yield down-gradient fluxes were identified along with the change in the up-gradient or
down-gradient character of the fluxes that occurs as the jets grow.  The third method was the development of a reduced dynamical system
that retains the fully nonlinear interactions in contrast to the \mbox{G--L} equation.  This reduced system is
based on an adiabatic assumption for the covariance changes and on a
Galerkin truncation of the dynamics retaining only the $n_c$ and $2n_c$ components of the
mean flow that play important role in the equilibration of the zonostrophic instability.

For the \mbox{G--L} branch, the central physical process responsible for the equilibration is the reduction in the up-gradient
vorticity flux that occurs through the change in the homogeneous part
of the eddy covariance.  For low values of $\b$, the instability
is quickly quenched and the jets equilibrate at low amplitude.  The reason is that the contribution of the eddies that induce up-gradient
fluxes and drive the instability is weakened as the jets emerge while simultaneously the contribution of the eddies that induce down--gradient
fluxes is increased.  As a result, the jets equilibrate at a small amplitude and are supported by the same eddies that drive the instability.

For
large values of $\b$, both the up-gradient and the down-gradient contributions are almost equally weakened thus leading to a slow decay of the
growth rate and to an equilibrated jet with a much larger amplitude.  Because the equilibrium amplitude is large, the  stabilizing fluxes that
are multiplied by the square of the jet amplitude in the \mbox{G--L} equation are dominant and, therefore, at equilibrium the jet is supported
by the eddies that were initially hindering its growth (these eddies have phase lines that form small angles with the meridional but
different than zero).

For the new branch of jet equilibria the main physical process responsible for the equilibration is the interaction
of the $n_c$ and the $2n_c$ component of the emerging flow.  Starting from a finite amplitude jet with either
strong $n_c$ or $2n_c$ components, this nonlinear interaction leads to rapid growth of the jet and to equilibration
of the flow at amplitudes much larger than the \mbox{G--L} branch and with much stronger $2n_c$ component.

Finally, the stability of the equilibrated side band unstable jet perturbations was examined.  For an infinite domain, zonal jets
with scales close to the scale $n_c$ of the most unstable mode of the flow-forming instability are stable; jets with scales much
larger or much smaller are unstable.  The incipient Eckhaus instability of the harmonic equilibria of the \mbox{G--L} equation
is well studied within the literature of pattern formation but here it was interpreted in a physically intuitive way.  The equilibrated jets have a low amplitude (proportional to the supercriticality) and therefore do not significantly change the structure of the turbulence.  As a result, a mean flow perturbation on the turbulent flow
induces approximately the same vorticity flux feedback as in the absence of any jet with the vorticity flux feedback having a maximum at the most unstable wavenumber.  Therefore, when a dilation--compression phase perturbation is inserted in the equilibrated jet that has a different wavenumber than $n_c$, the vorticity flux feedback for the dilated or the compressed part of the jet will be larger and this part of the jet tends to grow and take over the whole domain.

The predictions for the stability boundary and the growth rate of the Eckhaus instability were then compared to the stability analysis of the jet equilibria using the fully nonlinear S3T system and the methods developed in \citet{Constantinou-2015-phd}.  For $\beta\approx 5$, using the exact values for the curvature of the growth rate function yields accurate predictions for both the stability boundary and the
growth rate.  As the instability develops the unstable side band jets with smaller/larger scale than the jet with wavenumber $n_c$ branch/merge into the stable $n_c$ jet.  For low or high values of $\beta$, large quantitative discrepancies occur with a few exceptions,
but the qualitative picture of the dynamics with branching/merging into the stable jet equilibrium remains.

We note that the comparison of the G-L dynamics with nonlinear S3T integrations, as well as investigation of the equilibration process with an anisotropic ring forcing showed that the results in this study are not sensitive to the forcing structure.   

A question that rises naturally is whether the results discussed here are relevant for strong turbulent jets.
Strong turbulent jets also undergo bifurcations as the turbulence intensity increases. There are, however, qualitative differences compared to weak jets: strong jets \emph{always} merge to larger scales while weak jets can either merge or branch to reach a scale close to $n_c$. Based on the relevant dynamics in pattern formation, we expect that the anti-diffusive phase dynamics that are involved in the Eckhaus instability will play a significant role in the secondary instabilities of large-amplitude jets as well.  Moreover, the generalization of the Ginzburg--Landau dynamics that we have put forward in this study (eqs.~\eqref{eqs:u1u2tend}) is able to describe the slow evolution of a jet that consists of more than just one harmonic. This generalization of the Ginzburg--Landau dynamics, we hope, will provide a vehicle for understanding the dynamics involving bifurcations of strong turbulent jets.


\acknowledgments
The authors would like to thank Jeffrey B.  Parker for helpful comments on the first version of the manuscript.  N.A.B.~was supported by the AXA Research Fund.  N.C.C.~was partially supported by the NOAA Climate and Global Change Postdoctoral Fellowship Program, administered by UCAR's Cooperative Programs for the Advancement of Earth System Sciences and also by the National Science Foundation under Award OCE-1357047.

\ifdraft\else\clearpage\onecolumn\fi

\appendix[A]

\appendixtitle{S3T formulation and eigenvalue relation of the flow-forming instability}
\label{app:dispersion}
In this appendix we derive the eigenvalue relation of the flow-forming instability.  The eigenvalue relation was
first derived by~\citet{Srinivasan-Young-2012}.  Here, we repeat the derivation mainly to introduce some notation
and terminology that will prove to be helpful in understanding the nonlinear equilibration of the flow-forming
instability.

Consider the S3T system~\eqref{eq:s3ts}, where
\begin{align}
\Lcal\equiv\bit\beta\(\partial_{x_a}\Del_a^{-1}+\partial_{x_b}\Del_b^{-1}\) + 2 \ ,\ \label{eq:op_L}
\end{align}
is the operator governing the linear eddy dynamics,
\begin{align}
  \Ncal(\um,C)&\equiv\left[-\um_a\,\partial_{x_a}+(\partial_{y_a}^2\um_a)\,\partial_{x_a}\Del_a^{-1} -\um_b\,\partial_{x_b}+(\partial_{y_b}^2\um_b)\,\partial_{x_b}\Del_b^{-1}\right]C\ ,\  \label{eq:op_N}
\end{align}
is the nonlinear operator governing the eddy--mean flow interaction and
\begin{equation}
\Rcal(C) \equiv \frac{1}{2}\[\(\partial_{x_a}\Del_a^{-1}+\partial_{x_b}\Del_b^{-1}\)C\bit\]_{a=b}\ ,\label{eq:rs}
\end{equation}
is the eddy vorticity flux driving the mean flow.  Subscripts $a$ or $b$ on operators acting on $C$ indicate the point of
evaluation and the specific independent variable the operator is acting on, and the subscript $a=b$ indicates
that the function of $\bx_a$ and $\bx_b$, e.g., inside the square brackets on the right-hand-side of~\eqref{eq:rs}, is transformed
into a function of a single variable by setting $\bx_a=\bx_b=\bx$.

The eigenvalue relation is obtained by linearizing the S3T system~\eqref{eq:s3ts} about the homogeneous equilibrium~\eqref{eq:hom_eq}.
Then, introducing the ansantz~\eqref{eq:eigfunc} in the linearized S3T equations we obtain:
\begin{subequations}
\begin{align}
(\s +1)\d \um &= \Rcal(\d C) \label{eq:ds3tm}\ ,\\
(\s + \Lcal)\delta C &= \Ncal(\d\bar{u},C^e) \label{eq:dcov_evo}\ .
\end{align}
\end{subequations}
The quantity:
\begin{equation}
	f(\s|\delta \um, C)\equiv\Rcal\left[\left(\s+\Lcal\right)^{-1}\Ncal(\delta \um, C)\right]\ ,
\label{eq:rstress}
\end{equation}
is the  vorticity flux induced by the distortion of the incoherent homogeneous eddy equilibrium field with covariance $C$ by the mean flow $\d\um$.

The inversion of the operators and the algebra is simplified by taking the Fourier decomposition of $\tilde{C}^{(h)}_{n}$:
\begin{equation}
\tilde{C}^{(h)}_{n}(\bx_a-\bx_b)=\int \frac{\df^2\kv}{(2\pi)^2}\;\hat{C}(\kv)\,\ee^{\i\kv\bcdot(\bx_a-\bx_b)}\ .\label{eq:fourier1}
\end{equation}
By inserting~\eqref{eq:fourier1} and~\eqref{eq:hom_eq} into~\eqref{eq:dcov_evo} we obtain:
\begin{equation}
\delta C=\e\,\ee^{\i n(y_a+y_b)/2}\[G^{+}(\s,\bx_a-\bx_b)-G^{-}(\s|\bx_a-\bx_b)\]\ ,\label{eq:deltaC}
\end{equation}
where we defined
\begin{align}
   G^\pm(\s|\bx)\equiv\int\frac{\df^2\kv}{(2\pi)^2}\; \frac{\i k_x k_{\mp}^2(k_{\pm}^2- n ^2)}{(\s+2) k_{+}^2k_{-}^2+2\i\b  n  k_x k_y} \frac{\hat{Q}(\kv_{\pm})}{2}\ee^{\i\kv\bcdot\bx}\ ,\label{eq:G}
\end{align}
with $\kv_{\pm}=\kv+\nv/2$, $\nv = (0, n )$ and $k_\pm=|\kv_{\pm}|$.  Inserting~\eqref{eq:deltaC} in~\eqref{eq:ds3tm} we obtain~\eqref{eq:dispersion}, in which
\begin{equation}
f=\int\frac{\df^2\kv}{(2\pi)^2}\; \frac{ n  k_x^2(k_y+ n /2)\,(1- n ^2/k^2)}{(\s+2)
k^2k_{s}^2+2\i\b  n  k_x (k_y+ n /2)}\hat{Q}(\kv)\ ,\label{eq:ff}
\end{equation}
with $k_s \equiv |\kv+\nv|$.  After substituting the ring forcing power spectrum~\eqref{eq:Qhat},
expressing the integrand in polar coordinates $(k_x,k_y)=(k\cos\thet, k\sin\thet)$ and integrating
over $k$~\eqref{eq:ff} becomes:
\begin{equation}
f=\int_0^{2\pi}\frac{N_f\,\df\theta}{(\s+2)D_f+\i\b D_\beta}\ ,
\end{equation}
with $N_f(\theta)= n \cos^2\theta(\sin\thet+ n /2)(1- n ^2)/ \pi$, $D_f(\thet)=\cos^2\thet+(\sin\thet+ n )^2$ and
$D_\beta(\thet)=2 n \cos\thet(\sin\thet+  n /2)$.  At criticality ($\s=0$), using the mirror symmetry
property of the forcing, i.e.,~$\hat{Q}(-k_x,k_y)=\hat{Q}(k_x,k_y)$, the vorticity flux feedback is rewritten as:
\begin{equation}
f_r=\int_0^{\pi/2}\Fcal (\thet, n )\,\df\thet\ ,\label{eq:ff1}
\end{equation}
where
\begin{equation}
	\Fcal(\thet, n )=\frac{N_f(\thet)\,D_f(\thet)}{4D_f^2(\thet)+\b^2D_\beta^2(\thet)}+ \frac{N_f(\thet+\pi)\,D_f(\thet+\pi)}{4D_f^2(\thet+\pi)+\b^2D_\beta^2(\thet+\pi)}\ ,\label{eq:Fcal}
\end{equation}
is the contribution to the feedback from the waves with wavevectors $(k_x,k_y)$, $(-k_x,-k_y)$ and their mirror symmetric wavevectors
$(-k_x,k_y)$ and $(k_x,-k_y)$ respectively.

\appendix[B]

\appendixtitle{Ginzburg--Landau equation for the weakly nonlinear evolution of a zonal jet
perturbation about the homogeneous state\label{app:GL}}

To obtain the \mbox{G--L} equation governing the nonlinear S3T dynamics near the onset of the
instability, we assume that the energy input rate is slightly supercritical
$\e=\e_c(1+\mu^2)$, where $\mu\ll 1$ measures the supercriticality.  As
discussed in section~\ref{sec:equil}, the emerging jet grows slowly at a rate $O(\mu^2)$ and contains a
band of wavenumbers of $O(\mu)$ around~$n_c$, where $n_c$ is
the wavenumber of the jet that achieves neutrality at $\e_c$.  Therefore, we assume that
the dynamics evolve on a slow time scale $T=\mu^2t$ and are modulated at a long meridional scale $Y=\mu y$.  The leading order jet is $\um_1=A(Y,T)\,\ee^{\i n_c y}$.  We then expand the
velocity and the covariance as a series in $\mu$:\begin{subequations}\begin{align}
\um&= \mu\,\um_1(y,Y,T)+\mu^2\,\um_2(y,Y,T) + O(\mu^3)\ ,\\
C&= C^e(\bx_a-\bx_b)+\mu\,C_1(\bx_a-\bx_b,Y_a,Y_b,T)+\mu^2\,C_2(\bx_a-\bx_b,Y_a,Y_b,T) + O(\mu^3)\ ,
\end{align}\label{expand_mu}\end{subequations}
along with the linear and nonlinear operators $\Lcal$ and $\Ncal$ that depend on the fast and slow meridional coordinates, $y$ and~$Y$ respectively.  

We substitute~\eqref{expand_mu} in~\eqref{eq:s3ts} and collect terms with equal powers of $\mu$.  As discussed in section~\ref{sec:equil}, we further assume that the amplitude $A$, as well as $C_1$ and $C_2$, are independent of the slow coordinate~$Y$.  This way operators $\Lcal$ and $\Ncal$ also become independent of~$Y$.  In this case, the order $\mu^0$ terms yield the homogeneous equilibrium.  Terms of order $\mu^1$ yield
the balance:
\begin{equation}
\Acal\begin{pmatrix}\um_1\cr C_1\cr\end{pmatrix}\equiv \begin{pmatrix} \um_1-\Rcal(C_1)\cr \Lcal C_1-\Ncal(\um_1,C^e)\cr\end{pmatrix}=0\ ,
\end{equation}
which can also be compactly written as
\begin{equation}
\um_1=\e_c f\(0|\um_1,Q/2\)\ ,\label{eq:feedmu}
\end{equation}
where $f\(\s|\um_1,Q/2\)$ is the vorticity flux feedback on the mean flow $\um_1$ as defined in~\eqref{eq:rstress}.  The solution of~\eqref{eq:feedmu} is the eigenfunction of operator $\Acal$ with zero
eigenvalue:
\begin{equation}
\begin{pmatrix}\um_1\cr C_1\cr\end{pmatrix}=A(T)\begin{pmatrix}\ee^{\i n_c y}\cr \e_c \,\ee^{\i n_c (y_a+y_b)/2}
\left[\bit G_c^+(0|\bx_a-\bx_b)-G_c^-(0|\bx_a-\bx_b)\right]\cr\end{pmatrix}+\mbox{c.c.}\ .\label{eq:U1C1}
\end{equation}
In~\eqref{eq:U1C1} the subscript $c$ on $G^\pm$ denotes that they are evaluated at $n=n_c$.  At order $\mu^2$ the balance is:
\begin{equation}
\Acal\begin{pmatrix} \um_2\cr C_2\cr\end{pmatrix}=\begin{pmatrix} 0\cr \Ncal(\um_1,C_1)+\e_c Q\cr\end{pmatrix}\ .
\label{eq:ds3tsmu2}
\end{equation}
Equation~\eqref{eq:ds3tsmu2} has a homogeneous solution which is proportional to $[\um_1, C_1]^T$ and can be incorporated in it, and a particular solution.  The nonlinear term $\Ncal(\um_1,C_1)$ generates both a double and a zero harmonic mean flow (and covariance).  As a result, the particular solution is:
\begin{equation}
	\begin{pmatrix}\um_2\cr C_2\cr\end{pmatrix}=\begin{pmatrix}0\cr \e_c Q(\bx_a-\bx_b)/2+C_{20}(\bx_a-\bx_b,T)\cr\end{pmatrix} +\begin{pmatrix}\alpha_2\,A(T)^2\,\ee^{2\i  n_c y}\cr C_{22}(\bx_a-\bx_b,T)\,\ee^{2\i  n_c (y_a+y_b)/2}\cr\end{pmatrix}+\mbox{c.c.}\ ,\label{eq:U1C2}
\end{equation}
where $C_{20}$ and $C_{22}$ are the zero and double harmonic coefficients of the covariance and
\begin{equation}
	\alpha_2\defn\dfrac{\dfrac{\e_c}{2} \displaystyle\int\dfrac{\df^2\kv}{(2\pi)^2}\dfrac{\i n_c k_x^3 (k^2- n_c ^2) }{k^2 k^2_2 +\i\b n_c k_x k_{y,1}} \left\{\dfrac{k_{y,2} (k_2^2- n_c ^2)}{k^2k_4^2+2\i\b n_c k_x k_{y,2} }-\dfrac{ k_y k^2_2 (k^2- n_c ^2)} {k^2( k_{-2}^2 k_2^2+2\i\b n_c k_x k_y)}\right\}\hat{Q}(\kv)}{\e_c\displaystyle\int\dfrac{\df^2\kv}{(2\pi)^2}\dfrac{n_c k_x^2 k_{y,2} (k^2-4 n_c ^2)}{k^2( k^2 k_4^2+2\i\b n_c k_x k_{y,2})}\hat{Q}(\kv)-1}\ ,\label{eq:a1}
\end{equation}
with $k_{y,j}\equiv k_y + j n_c /2$ and $k_j^2\equiv k_x^2+k_{y,j}^2$ for any integer $j$.


At order $\mu^3$ the balance is:
\begin{equation}
\Acal\begin{pmatrix} \um_3\cr C_3\cr\end{pmatrix}=\begin{pmatrix} -\partial_T\um_1\cr
-\partial_TC_1+\Ncal(\um_2,C_1)+\Ncal(\um_1,C_2)\cr\end{pmatrix}\ .
\label{eq:ds3tsmu3}
\end{equation}
If the right-hand-side of~\eqref{eq:ds3tsmu3} is an eigenvector of operator $\Acal$ with zero eigenvalue then secular terms appear that produce a mean flow and an associated covariance that are unbounded at $|y|\to\infty$.  This occurs when
\begin{equation}
-\partial_T\um_1+\Rcal\left\{\Lcal^{-1}\left[ -\partial_TC_1+\Ncal(\um_2,C_1)+\Ncal(\um_1,C_2)\right]\right\}
\label{eq:sec1}
\end{equation}
has a non-zero $\ee^{\i n_c y}$ component.  The secular terms vanish if:
\begin{equation}
\partial_T \um_1+\Rcal\left(\Lcal^{-1}\partial_T C_1\right)= f(0\,|\,\um_1,C^e)+f(0\,|\,\um_1,C_{20}) +\Pcal_1 \left[f(0|\um_1,C_{22}\,\ee^{2\i  n_c (y_a+y_b)/2}+\mbox{c.c.})+f(0|\um_2,C_1)\right]\ .\label{eq:GL1}
\end{equation}
where $\Pcal_1$ is the operator that projects onto the harmonic~$n_c$:
\begin{equation}
\Pcal_1 g(y)\equiv \dashint g(s)\,\ee^{\i n_c (y-s)}\df s\ .
\end{equation}

Equation~\eqref{eq:GL1} determines the equilibration of the most unstable jet.  The terms on the right-hand-side
of~\eqref{eq:GL1} are nonlinear in $\um$ and $C$ and they are responsible for the equilibration of the SSD instability. Let us take a closer look into each term in~\eqref{eq:GL1}.  The second term on the left-hand-side of~\eqref{eq:GL1} is:
\begin{equation}
\Rcal\left(\Lcal^{-1}\partial_TC_1\right)=\(\partial_T A\)(c_1-1)\,\ee^{\i  n_c y}\ ,\label{eqB1}
\end{equation}
where
\begin{equation}
c_1=1+\frac{\e_c}{4}\int\frac{\df^2\kv}{(2\pi)^2}\frac{ n_c k_x^2k_{y,1} k_2^2 (k^2- n_c ^2)}
{(k^2 k^2_2 +\i\b n_c k_x k_{y,1})^2}\hat{Q}(\kv)\ .\label{eq:c1}
\end{equation}
The first term on the right-hand-side of~\eqref{eq:GL1} is the vorticity flux feedback
on $\um_1$ at criticality
\begin{equation}
f(0|\um_1,C^e)=A\,\ee^{\i n_c y}\ .\label{eqB2}
\end{equation}
The second term on the right-hand-side of~\eqref{eq:GL1} is the
vorticity flux feedback between the order $\mu^1$ mean jet $\um_1$, and the homogeneous order $\mu^2$ eddy covariance $C_{20}$:
\begin{equation}
f(0|\um_1,C_{20})=-c_3^{ec}A|A|^2\,\ee^{\i n_c y}\ ,\label{eqB3}
\end{equation}
with\ifdraft{\small\else\fi
\begin{equation}
  c_3^{ec}\defn\frac{\e_c}{4}\int\frac{\df^2\kv}{(2\pi)^2}\frac{ n_c  k_x^4k_2^2(k_2^2- n_c ^2)(k^2- n_c ^2)^2} {|k^2 k^2_2 +\i\b n_c k_x k_{y,1}|^2}\left[\frac{2k_{y,1}}{
k^2 k^2_2 +\i\b n_c k_x k_{y,1}}-\frac{k_{y,-1}}{k^2 k_{-2}^2 +\i\b n_c k_x k_{y,-1}}-\frac{k_{y,3}}{k_2^2 k_4^2
+\i\b n_c  k_x k_{y,3}}\right]\hat{Q}(\kv)\ .\label{eq:c3_ec}
\end{equation}\ifdraft}\else\fi
The third term on the right-hand-side of~\eqref{eq:GL1} is the $\ee^{\i n_cy}$ component of the vorticity flux feedback
between the jet $\um_1$, with wavenumber $n_c$ and the jet $\um_2$ with wavenumber $2n_c$ with the inhomogeneous
eddy covariance $C_1$ and $C_{22}$:
\begin{equation}
\Pcal_1\left[f(0|\um_1,C_{22}\,\ee^{2\i  n_c (y_a+y_b)/2}+\mbox{c.c.})+f(0|\um_2,C_1)\right]=-c_3^{1,2}A|A|^2\,\ee^{\i  n_c y}\ ,\label{eqB4}
\end{equation}
with\ifdraft{\normalsize\else\fi
\begin{align}
c_3^{1,2}&\defn\frac{\e_c}{8}\int\frac{\df^2\kv}{(2\pi)^2}\left\{\vphantom{\sum^{\bit}_{\bit}}\right.
\frac{ n_c  k_x^4k^2(k^2- n_c ^2)(k_2^2- n_c ^2)}
{\left[k^2 k^2_4 +2\i\b n_c k_x k_{y,2}\right]\left[k^2 k^2_2 +\i\b n_c k_x k_{y,1}\right]}
\left[\frac{k_{y,1} (k_4^2- n_c ^2)}{k^2 k^2_2 +\i\b n_c k_x k_{y,1}}-\frac{k_{y,3}k_4^2(k^2- n_c ^2)}
{k^2(k^2_2 k^2_4 +\i\b n_c k_x k_{y,3})}\right]\nonumber\\
&\hspace{5em}+\frac{n_c  k_x^4 k_{y,1} k_2^2(k_{-2}^2- n_c ^2)(k^2- n_c ^2)^2 (k_{-2}^2 k_2^2+\i\b n_c k_x k_y) }{(k^2 k^2_2 +\i\b n_c k_x k_{y,1})^2 (k^2 k_{-2}^2 +\i\b n_c k_x k_{y,-1} )(k_{-2}^2k_2^2+2\i\b n_c k_x k_y)}
\left.\vphantom{\sum^{\bit}_{\bit}}\right\}\hat{Q}(\kv)\nonumber\\
&+\i\,a_2\,\frac{\e_c}{4}\int\frac{\df^2\kv}{(2\pi)^2}\left\{\vphantom{\sum^{\bit}_{\bit}}\right.
\frac{k^2-4 n_c ^2}{k^2 k^2_4 +2\i\b n_c k_x k_{y,2}}\left[\frac{k_{y,1} (k_4^2- n_c ^2)}{k^2 k^2_2 +\i\b n_c k_x k_{y,1}}-\frac{k_{y,3}k_4^2(k^2- n_c ^2)}
{k^2(k^2_2 k^2_4 +\i\b n_c k_x k_{y,3})}\right]\nonumber\\
&\hspace{7em}+\frac{k^2- n_c ^2}{k^2k_2^2-\i\b n_c k_x k_{y,1}}\left[\frac{k_{y,-1} (k_2^2-4 n_c ^2)}
{k^2k_{-2}^2 +\i\b n_c k_x k_{y,-1} }-\frac{ k_{y,3} k_2^2(k^2-4 n_c ^2)}{k^2(k^2_2 k^2_4 +\i\b n_c k_x k_{y,3})}\right]
\left.\vphantom{\sum^{\bit}_{\bit}}\right\} n_c k_x^3\hat{Q}(\kv)\ .\label{eq:c3_12}
\end{align}\ifdraft}\else\fi
Therefore, using~\eqref{eqB1},~\eqref{eqB2},~\eqref{eqB3} and~\eqref{eqB4} we get that~\eqref{eq:GL1} reduces to:
\begin{equation}
  c_1\,\partial_T A=A-c_3\,A|A|^2\ ,\label{eq:GL2}
\end{equation}
where $c_3\defn c_3^{ec}+c_3^{1,2}$.

Finally, we arrive to the \mbox{G--L} equation~\eqref{eq:GL} by adding the diffusion term $c_2\partial_{Y}^2A$ on the~right-hand-side of~\eqref{eq:GL2}, with
\ifdraft{\normalsize\else\fi\begin{align}
	&c_2\defn-\frac{\e_c}{2}\left(\frac{\partial^2 f}{\partial  n ^2}\right)_{ n_c ,\s=0}\nonumber\\
	&=\frac{\e_c}{2}\int\frac{\df^2\kv}{(2\pi)^2}\Bigg[
	\frac{k_x^2 k_{y,2}^2 k^2(k^2- n_c ^2)(2k^2+\i \b k_x )}{(k^2 k^2_2 +\i\b n_c k_x k_{y,1})^3}-\frac{k_x^2 k^2(k^2-4 n_c k_y-5 n_c ^2)}{2(k^2 k^2_2 +\i\b n_c k_x k_{y,1})^2}+\frac{ n_c k_x^2 k_{y,1} }{k^2(k^2 k^2_2 +\i\b n_c k_x k_{y,1})} \Bigg]\hat{Q}(\kv)\ .\label{eq:c2}
\end{align}\ifdraft}\else\fi
The coefficients $c_1$,~$c_2$ and $c_3$ are all functions of $\b$, $n_c$ and the forcing covariance spectrum, $\hat{Q}$.
For the ring forcing~\eqref{eq:Qhat} considered here they are all real and positive.  

\begin{figure}
\centerline{\includegraphics[width=19pc]{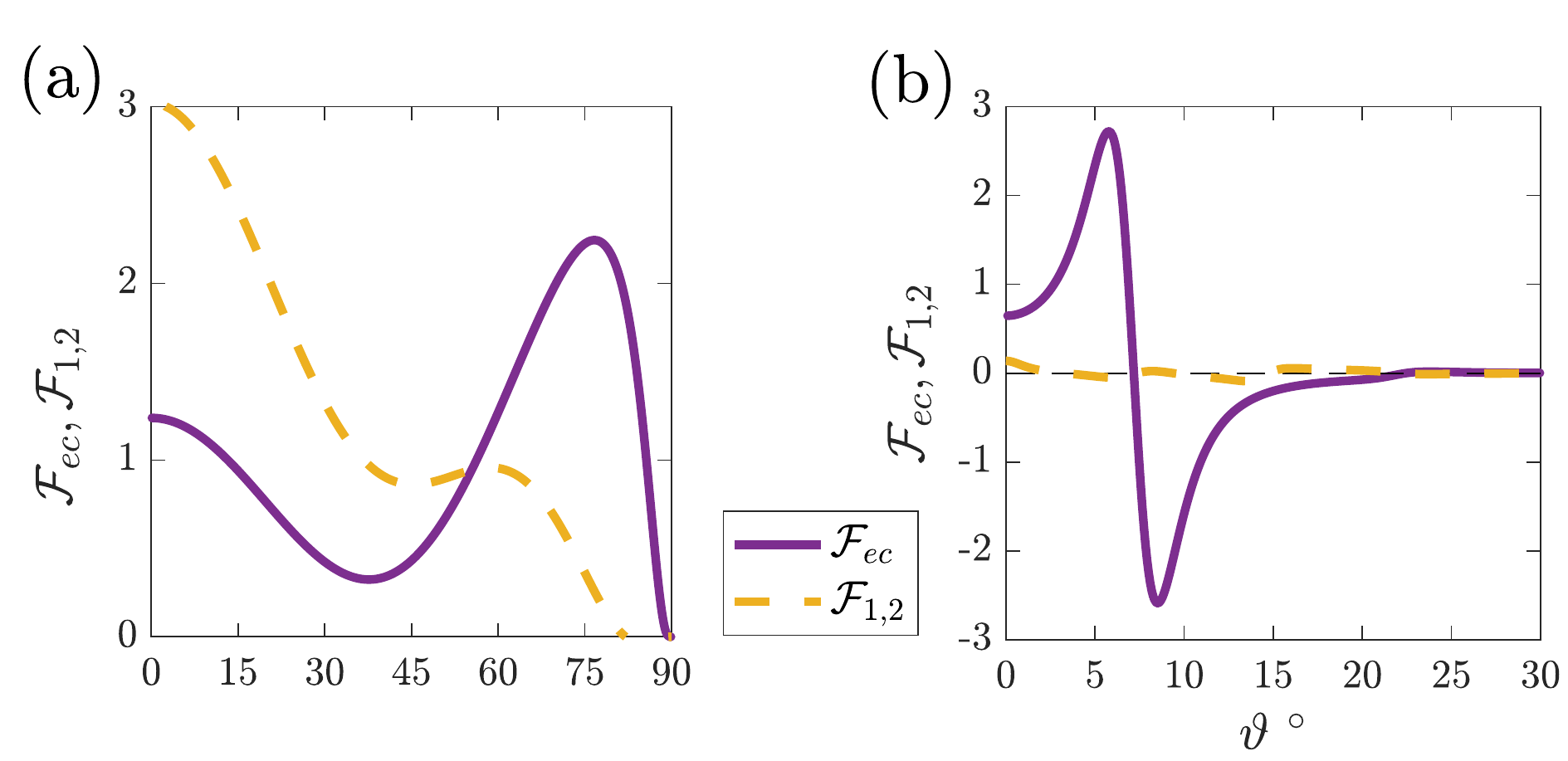}}
\vspace{-1em}
\caption{The contribution of the two feedbacks $\Fcal_{ec}$ (solid) and $\Fcal_{1,2}$ (dashed) to the nonlinear coefficient $\Fcal_{\rm NL}$.  Panel~(a) shows the case with $\beta=0.1$ while panel~(b) with $\beta=100$.} \label{fig:flux_feedback_b}
\end{figure}

To study the contribution to each of the components of $c_3$ from the forced waves with phase lines forming an
angle $\vartheta$ with the $y$-axis, we substitute the ring forcing power spectrum~\eqref{eq:Qhat}.
After expressing the integrand in polar coordinates $(k_x,k_y)=(k\cos\thet, k\sin\thet)$ and integrate over $k$ we obtain:
\begin{equation}
\left[\vphantom{c_3^{ec}}\right.  c_3^{ec}, c_3^{1,2}, c_3\left.\vphantom{c_3^{ec}}\right]=\e_c\int_0^{\pi/2}
\left[\Fcal_{ec}, \Fcal_{1,2}, \Fcal_{\rm NL}\right]\,\df\thet\  ,\label{eq:F_NL}
\end{equation}
where $\Fcal_{ec}$, $\Fcal_{1,2}$, and $\Fcal_{\rm NL}$ is the contribution of
the waves with $(k_x,k_y)$, $(-k_x,-k_y)$ and their mirror symmetric $(-k_x,k_y)$ and $(k_x,-k_y)$ to the feedbacks and
$\Fcal_{\rm NL}=\Fcal_{ec}+\Fcal_{1,2}$.  \ifdraft Figure~B10 \else Figure~\ref{fig:flux_feedback_b} \fi shows these contributions
as a function of wave angle.  For $\beta\ll 1$, forced eddies at all angles contribute positively to both
$\Fcal_{ec}$ and $\Fcal_{1,2}$.  The eddies tend to reduce the positive destabilizing contribution $\Fcal>0$ at small angles mainly
through $\Fcal_{1,2}$, while they enhance the negative stabilizing contribution $\Fcal<0$ at large angles mainly through $\Fcal_{ec}$.
For $\beta\gg1$, the dominant contribution comes from $\Fcal_{ec}$ and it follows roughly the same pattern as $\Fcal$.  That is, due to
the reduction in their energy the eddies tend to reduce both the up-gradient vorticity fluxes of waves with angles $|\thet| \lessapprox \thet_0$
and the down-gradient fluxes of waves with phase lines at angles $|\thet|\gtrapprox \thet_0$ with the latter reduction being larger.  As a
result, the nonlinear feedback of eddies with phase lines at angles $|\thet|\gtrapprox \thet_0$ is to enhance the jet and, as discussed
in section~\ref{sec:equil}, these are the eddies that support the equilibrated jet.

\appendix[C]
\appendixtitle{Non-isotropic ring forcing\label{app:NIF}}


Here we briefly discuss the effect of the forcing anisotropy on the obtained results. Consider the generalization of
forcing~\eqref{eq:Qhat} with spectrum:
\begin{equation}
	\Qhatd(\kvd) = 4\pi\,\kfd\,\d(\kd-\kfd)\,\left[1 + \gamma \cos (2 \thet )\right]\ ,\label{eq:Qhatgen}
\end{equation}
where $\thet\defn\arctan(\kyd/\kxd)$ and $|\gamma| \le1$ so that $\Qhatd \ge 0$. Parameter $\gamma$ determines the degree of anisotropy of the forcing~\citep{Srinivasan-Young-2014,Bakas-etal-2015}. The isotropic case of~\eqref{eq:Qhat} is recovered for $\gamma=0$.  For example, for $\gamma=1$ we get an anisotropic forcing that favors structures with small $|\kyd|$ (i.e., favoring structures like that in Fig.~\ref{fig:forcing}(a)
compared to structures like that in Fig.~\ref{fig:forcing}(b)), as if the vorticity injection was due to baroclinic growth processes. All three coefficients $c_1$, $c_2$, and~$c_3$ in~\eqref{eq:GL} are real and positive for forcing~\eqref{eq:Qhatgen}.

We first note that we obtain similar results to the isotropic forcing case regarding the comparison of the G-L predictions to the fully nonlinear dynamics (not shown). That is, both the existence of the upper branch equilibria, as well as the relative quantitative success of the G-L dynamics (after the proposed modifications) in predicting the amplitude and instability of the equilibrated jets are insensitive to forcing structure.

\begin{figure}
\centerline{\includegraphics[width=19pc]{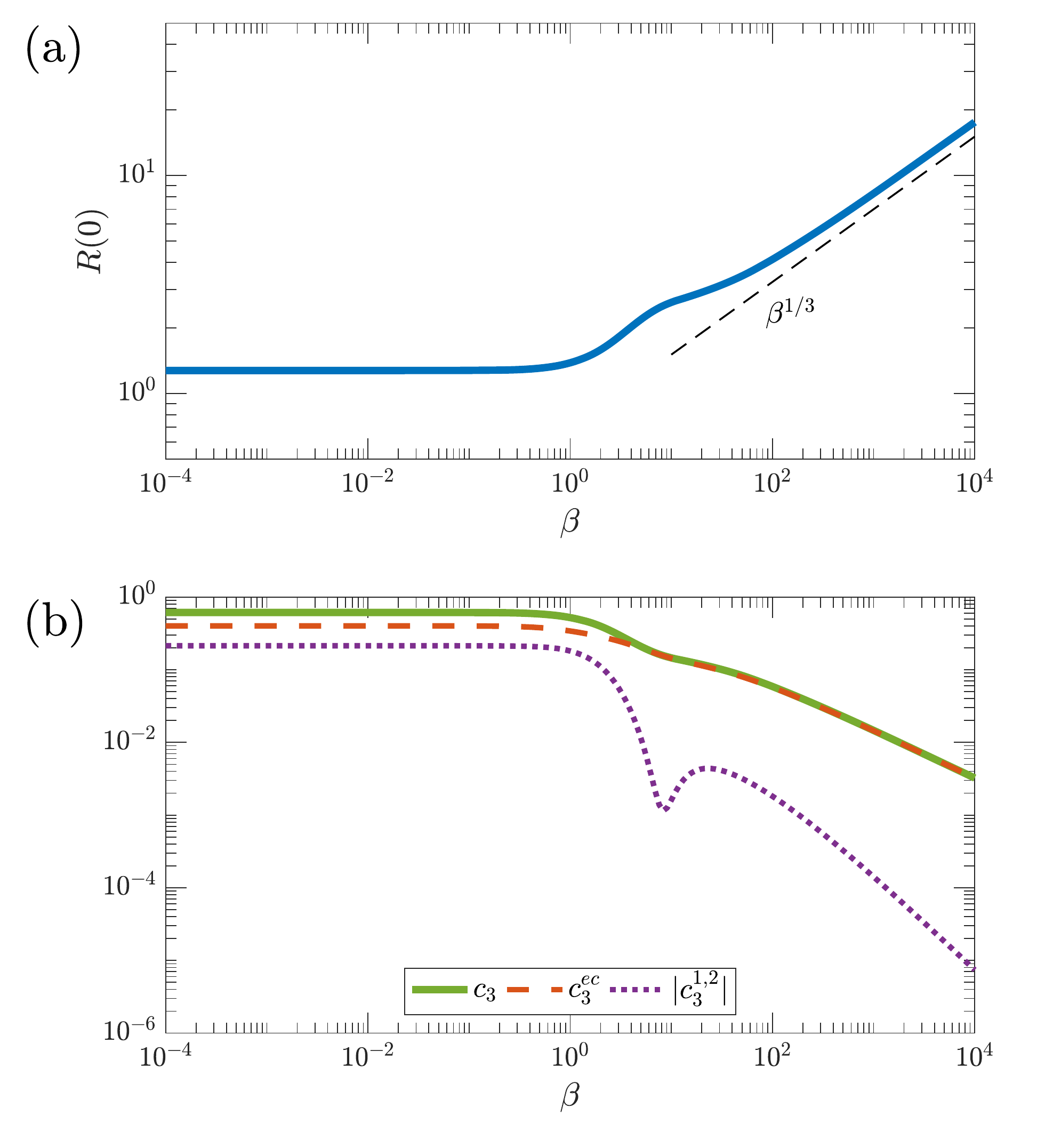}}
\vspace{-1em}
\caption{Same as Fig.~\ref{fig:R0_c3_gamma0} but for anisotropic forcing with $\gamma=1$.  (a) The amplitude $R_0(0) = 1/\sqrt{c_3}$ of the equilibrated most unstable jet with wavenumber $n_c$ as a function of $\beta$ for the case with  isotropic ($\gamma=0$) and anisotropic ($\gamma=1$) forcing.  Dashed line show the $\beta^{1/3}$ slope for reference.  (b) The coefficient $c_3$ and its decomposition into the contributions $c_3^{ec}$ and $c_3^{1,2}$ as a function of $\beta$.}
\label{fig:R0_c3_gamma1}
\end{figure}

Regarding the physical processes underlying the equilibration of the jets, we show in Fig.~\ref{fig:R0_c3_gamma1}(a) the amplitude $R_0$ for the equilibrated most unstable jet as a function of $\beta$. For $\beta\gg1$, the amplitude has the same power law as in the isotropic forcing case shown in Fig.~\ref{fig:R0_c3_gamma0}(a). However, the amplitude shows different dependence with $\beta$ for $\b\ll1$ but, however, this regime is of no interest since for as $\beta\rightarrow 0$ no zonal jets emerge in \eqref{eq:NLbarotropic} anyway. The relative contribution of the eddy-correction term and the interaction of $n_c$ with the double harmonic jet in $c_3$ is shown in Fig.~\ref{fig:R0_c3_gamma1}(b). Similarly to the isotropic forcing case, for most values of $\beta$ the equilibration is dominated by the interaction of the most unstable jet with the homogeneous covariance correction.

Lastly, we note that for anisotropic forcing similar qualitative decomposition of $c_3$ from various waves (as in Fig.~\ref{fig:FNL_betaF_R2FNL_beta}) also occurs  (not shown).

\appendix[D]
\appendixtitle{Eckhaus stability of \mbox{G--L} dynamics\label{app:GLstab}}

To address the Eckhaus instability of the harmonic
jet equilibria, we rewrite the jet amplitude $A$ in polar form
\eqref{def_A_thet}, we then substitute into~\eqref{eq:GL} and
separate real and imaginary parts to obtain:\begin{subequations}\begin{align}
c_1\partial_TR&=\left[1+c_2\partial_{Y}^2-c_2(\partial_Y\Theta)^2\right]R-c_3R^3\ ,\\
c_1R\partial_T\Theta&=2c_2(\partial_YR)(\partial_Y\Theta)+c_2R\partial_{Y}^2\Theta\ .
\end{align}\label{eq:GL_Rphi}\end{subequations}

Assume now an equilibrium jet with constant amplitude $R_0(\nu)$ and a linearly varying
phase $\Theta=\nu Y$.  Consider small perturbations about this equilibrium jet:
\begin{equation}
 R=R_0(\nu)+\rho\quad\text{and}\quad \Theta=\nu Y+\phi\ ,\label{eq46}
\end{equation}
and linearize~\eqref{eq:GL_Rphi} to obtain:\begin{subequations}\begin{align}
c_1\partial_T\rho&=\left[1+c_2(\partial_{Y}^2-\nu^2)-3c_3R_0^2\right]\rho-2c_2R_0\nu\partial_Y\phi\ ,\\
c_1R_0\partial_T\phi&=2c_2\nu\partial_Y\rho+c_2R_0\partial_{Y}^2\phi\ .
\end{align}\label{eq:Eckhaus_lin}\end{subequations}
Using the ansatz $[\rho,\phi]=[\hat{\rho},\hat{\phi}]\,\ee^{\i qY+\lambda T}$ we find that the eigenvalues $\lambda$ are:
\begin{equation}
\lambda=\frac{\nu^2-\nu_e^2-q^2\pm\sqrt{(\nu^2-\nu_e^2)^2+4q^2\nu^2}}{c_1\nu_e^2}\ .\label{eq:Eck_growth}
\end{equation}
Instability occurs when $\lambda> 0$, that is when
\begin{equation}
q^2+2(\nu_e^2-3\nu^2)< 0.\label{eq:Eck_stab_bound}
\end{equation}

\ifdraft\else\twocolumn\fi

\end{document}